\documentclass[useAMS, usenatbib, usegraphicx]{mn2e}
\pdfoutput=1

\usepackage[fleqn]{amsmath} 
\usepackage{amsfonts} 
\usepackage{amstext}
\usepackage{amssymb} 
\usepackage[colorlinks=true, citecolor=blue, linkcolor=blue, breaklinks=true]{hyperref}
\usepackage{subfig}

\newcommand{\hi } {{\rm H}\,{\small\rm I} \,}
\newcommand{\hiA } {{\rm H}\,{\small\rm I}}
\newcommand{\hii } {{\rm H}\,{\small\rm II} \,}

\title[The triggering of starbursts in low-mass galaxies]{The triggering of starbursts in low-mass galaxies}
\author[Lelli, Verheijen, $\&$ Fraternali]{Federico Lelli$^{1, 2}$ \thanks{E-mail: federico.lelli@case.edu},
Marc Verheijen$^{2}$, Filippo Fraternali$^{3, 1}$\\
$^{1}$Department of Astronomy, Case Western Reserve University,
10900 Euclid Ave, Cleveland, OH 44106, USA \\
$^{2}$Kapteyn Astronomical Institute, University of Groningen,
Postbus 800, 9700 AV, Groningen, The Netherlands \\
$^{3}$Department of Physics and Astronomy, University of Bologna,
via Berti Pichat 6/2, 40127, Bologna, Italy}

\begin{document}

\date{}

\maketitle

\begin{abstract}
Strong bursts of star formation in galaxies may be triggered either
by internal or external mechanisms. We study the distribution and
kinematics of the \hi gas in the outer regions of 18 nearby starburst
dwarf galaxies, that have accurate star-formation histories from HST
observations of resolved stellar populations. We find that starburst
dwarfs show a variety of \hi morphologies, ranging from heavily disturbed
\hi distributions with major asymmetries, long filaments, and/or
\hiA-stellar offsets, to lopsided \hi distributions with minor
asymmetries. We quantify the outer \hi asymmetry for both our
sample and a control sample of typical dwarf irregulars. Starburst
dwarfs have more asymmetric outer \hi morphologies than typical
irregulars, suggesting that some external mechanism triggered the
starburst. Moreover, galaxies hosting an old burst ($\gtrsim$100 Myr)
have more symmetric \hi morphologies than galaxies hosting a young one
($\lesssim$100 Myr), indicating that the former ones probably had enough
time to regularize their outer \hi distribution since the onset of the
burst. We also investigate the nearby environment of these starburst
dwarfs and find that most of them ($\sim$80$\%$) have at least one
potential perturber at a projected distance $\lesssim$200~kpc. Our results
suggest that the starburst is triggered either by past interactions/mergers
between gas-rich dwarfs or by direct gas infall from the IGM.
\end{abstract}


\section{Introduction}\label{sec:intro}

The mechanisms that trigger strong bursts of star-formation in
low-mass galaxies are poorly understood. Unlike spiral galaxies,
gas-rich dwarfs usually do not have density waves and stellar
bars, thus internal mechanisms such as bar-driven gas inflows
are generally ruled out \citep[e.g.][]{Hunter2004}. Other
internal mechanisms have been proposed, like torques due to
massive star-forming clumps \citep{Elmegreen2012}, triaxial
dark matter haloes \citep{Bekki2002}, or bars made of dark
matter \citep{Hunter2004}. External mechanisms are also possible,
like tidal perturbations from nearby companions \citep[e.g.]
[]{Noguchi1988}, interactions/mergers between gas-rich dwarfs
\citep[e.g.][]{Bekki2008}, or cold gas accretion from the IGM
\citep[e.g.][]{Silk1987}. In particular, cosmological models
predict that low-mass galaxies should accrete most of their
gas through cold flows, reaching the central parts of the
dark matter halo without being shock-heated to the virial
temperature \citep[e.g.][]{Dekel2006}. This process may still
take place at $z\simeq0$ in low-density environments \citep[]
[]{Keres2005}, thus isolated starburst dwarfs in the nearby
Universe are prime locations to search for cold gas accretion.

In the literature, starburst dwarfs are referred to with
several names, often related to the observational technique
used to identify the starburst. Common names are (i) blue
compact dwarfs (BCDs) as they have blue colors and high surface
brightnesses \citep[e.g.][]{GilDePaz2003}; (ii) \hii galaxies
as they have integrated spectra with strong emission lines
\citep[e.g.][]{Terlevich1991, Taylor1995}; and (iii) amorphous
dwarfs as they may have peculiar morphologies dominated by a
few giant star-forming regions \citep[e.g.][]{Gallagher1987,
Marlowe1999}. Hereafter we use the general term ``starburst
dwarfs'' to indicate any low-mass galaxy experiencing an
enhanced period of star-formation activity.

\begin{table*}
\begin{minipage}{\textwidth}
\caption{Galaxy Sample.}
\small
\setlength{\tabcolsep}{4pt}
\begin{center}
\begin{tabular}{@{}l l c c c c c c c c c}
\hline
Name     & Alternative  & R.A.       & Dec.      & $V_{\rm{sys}}$ & $V_{\rm rot}$        & Dist & $M_{\rm{R}}$   & $R_{\rm{opt}}$ & 12+log(O/H) & Ref. \\
         & Name         & (J2000)    & (J2000)   &\multicolumn{2}{c}{-- (km s$^{-1}$) --}& (Mpc) & (mag)          & (kpc)&          & \\
\hline
NGC 625  & ESO 297-G005 & 01 35 04.3 & -41 26 15 & 398$\pm$6 & 30$\pm$5 & 3.9$\pm$0.4 & -17.25$\pm$0.24 & 3.3 & 8.08$\pm$0.12& a, g, l  \\
NGC 1569 & UGC 3056     & 04 30 49.0 & +64 50 53 & -80$\pm$10& 50$\pm$5 & 3.4$\pm$0.2 & -17.14$\pm$0.25 & 3.0 & 8.19$\pm$0.02& a, h, m  \\
NGC 1705 & ESO 158-G013 & 04 54 13.9 & -53 21 25 & 635$\pm$2 & 72$\pm$10& 5.1$\pm$0.6 & -16.35$\pm$0.26 & 1.5 & 8.21$\pm$0.05& b, i, l \\
NGC 2366 & UGC 3851     & 07 28 51.9 & +69 12 34 & 103$\pm$1 & 49$\pm$6 & 3.2$\pm$0.4 & -16.64$\pm$0.27 & 4.4 & 7.91$\pm$0.05& a, h, l  \\
NGC 4068 & UGC 7047     & 12 04 02.7 & +52 35 28 & 206$\pm$2 & 39$\pm$5 & 4.3$\pm$0.1 & -15.67$\pm$0.05 & 1.8 & ...          & a, h \\
NGC 4163 & UGC 7199     & 12 12 09.0 & +36 10 11 & 158$\pm$4 & 10$\pm$5 & 3.0$\pm$0.1 & -14.81$\pm$0.10 & 1.0 & 7.56$\pm$0.14& a, h, l \\
NGC 4214 & UGC 7278     & 12 15 38.8 & +36 19 39 & 291$\pm$1 & 81$\pm$9 & 2.7$\pm$0.2 & -17.77$\pm$0.24 & 2.2 & 8.22$\pm$0.05& a, h, l  \\
NGC 4449 & UGC 7592     & 12 28 10.8 & +44 05 37 & 210$\pm$5 & 35$\pm$5 & 4.2$\pm$0.5 & -18.88$\pm$0.26 & 3.3 & 8.26$\pm$0.09& a, h, l  \\
NGC 5253 & Haro 10      & 13 39 56.0 & -31 38 31 & 410$\pm$10& ...      & 3.5$\pm$0.4 & -17.61$\pm$0.27 & 2.1 & 8.12$\pm$0.05& a, g, k \\
NGC 6789 & UGC 11425    & 19 16 41.9 & +63 58 17 & -151$\pm$2& 57$\pm$9 & 3.6$\pm$0.2 & -15.09$\pm$0.14 & 0.7 & ...          & a, i \\
UGC 4483 & ...          & 08 37 03.4 & +69 46 31 & 158$\pm$2 & 19$\pm$2 & 3.2$\pm$0.2 & -12.97$\pm$0.19 & 0.6 & 7.56$\pm$0.03& a, i, l \\
UGC 6456 & VII Zw 403   & 11 27 57.2 & +78 59 48 &-102$\pm$4 & 10$\pm$5 & 4.3$\pm$0.1 & -14.41$\pm$0.05 & 1.2 & 7.69$\pm$0.01& a, j, n \\
UGC 6541 & Mrk 178      & 11 33 28.9 & +49 14 22 & 250$\pm$2 & ...      & 4.2$\pm$0.2 & -14.61$\pm$0.10 & 0.9 & 7.82$\pm$0.06& c, j, l \\
UGC 9128 & DDO 187      & 14 15 56.8 & +23 03 22 & 150$\pm$3 & 24$\pm$4 & 2.2$\pm$0.1 & -12.82$\pm$0.12 & 0.6 & 7.75$\pm$0.05& a, h, l \\
UGCA 290 & Arp 211      & 12 37 22.1 & +38 44 41 & 468$\pm$5 & ...      & 6.7$\pm$0.4 & -14.09$\pm$0.18 & 0.9 & ...          & d, i \\
I Zw 18  & Mrk 116      & 09 34 02.0 & +55 14 25 & 767$\pm$4 & 38$\pm$4 &18.2$\pm$1.4 & -14.99$\pm$0.26 & 0.5 & 7.20$\pm$0.01& e, j, k \\
I Zw 36  & Mrk 209      & 12 26 16.8 & +48 29 39 & 277$\pm$2 & 29$\pm$2 & 5.8$\pm$0.5 & -14.88$\pm$0.23 & 0.9 & 7.77$\pm$0.01& f, i, k \\
SBS 1415+437 & ...      & 14 17 02.1 & +43 30 19 & 616$\pm$2 & 18$\pm$2 &13.6$\pm$1.4 & -15.90$\pm$0.25 & 2.4 & 7.62$\pm$0.03& a, i, o \\
\hline
\end{tabular}
\end{center}

\medskip
\textbf{Notes.} The galaxy center is derived from $R$-band or $V$-band images, while the rotation
velocity $V_{\rm rot}$ is measured at the outermost radius accessible by \hi data at relatively high
spatial resolutions (LVF14). Distances are derived from the tip of the red giant branch (TRGB).
The optical radius $R_{\rm opt}$ is defined as 3.2 exponential scale-lengths. The last column
provides references for the distance, the integrated photometry, and the ionized gas metallicity,
respectively.

\medskip
\textbf{References.} (a)~\citet{McQuinn2010a}; (b)~\citet{Annibali2003};
(c)~\citet{SchulteLadbeck2000}; (d)~\citet{Crone2002}; (e)~\citet{Annibali2013};
(f)~\citet{SchulteLadbeck2001}; (g)~\citet{Lauberts1989}; (h)~\citet{Swaters2002b};
(i)~\citet{GilDePaz2003}; (j)~\citet{Papaderos2002}; (k)~\citet{Izotov1999};
(l)~\citet{Berg2012}; (m)~\citet{Kobulnicky1997}; (n)~\citet{Thuan2005}; (o)~\citet{Guseva2003}.
\label{tab:sample}
\end{minipage}
\end{table*}

In \citet[][hereafter LVF14]{Lelli2014b}, we studied the \hi content
of 18 starburst dwarfs and found that disturbed \hi kinematics
are more common in starburst dwarfs ($\sim$50$\%$) than in typical
star-forming irregulars (Irrs, $\sim$10$\%$). This may be
related to the starburst trigger (interactions/mergers or disk
instabilities), but may also be a consequence of feedback from
supernovae and stellar winds, making it difficult to distinguish
between different triggering mechanisms. About 50$\%$ of our
starburst dwarfs, instead, have a regularly rotating \hi disk
that allows us to derive rotation curves and investigate the
internal mass distribution. In \citet{Lelli2014a}, we found
that the inner rotation curves of starburst dwarfs rise more
steeply than those of typical Irrs \citep[see also][]{vanZee2001,
Lelli2012a, Lelli2012b}, suggesting that there is a close link
between the intense star-formation activity and the shape of
the gravitational potential. A central concentration of mass
(gas, stars, and dark matter) seems to be a characterizing
property of starburst dwarfs and must be tightly linked to
the mechanism (either internal or external) that triggers
the starburst \citep[see][for an in-depth discussion]{Lelli2014a}.

Environmental studies also provide important clues about the
triggering mechanism. In general, star-forming dwarfs tend
to populate low-density environments \citep[e.g.][]{Iovino1988,
Salzer1989, Telles2000, Lee2000} and are not necessarily associated
with \textit{bright} galaxies \citep[e.g.][]{CamposAguilar1991,
CamposAguilar1993, Telles1995, Pustilnik1995}, suggesting that
tidal interactions with \textit{massive} companions are \textit{not}
the dominant starburst trigger. The possibility of interactions
with low-luminosity, low-surface-brightness (LSB) galaxies,
however, remains open \citep[e.g.][]{Mendez1999, Mendez2000,
Noeske2001, Pustilnik2001}, given that these objects are
usually under-represented in optical catalogs. Moreover, if
the starburst is due to a \textit{past} interaction/merger
with a LSB dwarf, the resulting tidal features would have
very low surface brightnesses and be difficult to unambiguously
identify, unless deep optical imaging is available \citep[e.g.]
[]{LopezSanchez2010a, Delgado2012}.

Alternatively, deep 21~cm-line observations can be used to search
for gas-rich companions, infalling gas, or signatures of past
interactions/mergers \citep[e.g.][]{Sancisi2008}. \citet{Taylor1993,
Taylor1995, Taylor1996} obtained low-resolution \textit{Very Large
Array} (VLA) observations of 21 starburst dwarfs and 17 LSB dwarfs,
and concluded that starburst dwarfs have nearby ``\hi companions''
more than twice as often as LSB dwarfs. However, given the poor
angular resolution of these observations, in most cases it was
not possible to distinguish between actual \hi companions and
asymmetries in the outer \hi distribution. Detailed studies
of the \hi gas in the outer regions of starburst dwarfs have
been so far restricted to individual objects or small galaxy
samples. These works have revealed that several starburst dwarfs
show extended and filamentary \hi structures, which may indicate
either a recent interaction/merger or cold gas accretion from
the environment; e.g. NGC~1569 \citep{Stil2002}, IC~10
\citep{Manthey2008}, and NGC~5253 \citep{LopezSanchez2012}.
In some cases, the presence of a nearby companion and/or of
stellar tidal features clearly points to an interaction/merger
between gas-rich dwarfs; e.g. II~Zw~40 \citep{vanZee1998b},
II~Zw~70/71 \citep{Cox2001}, and I~Zw~18 \citep{Lelli2012a}.
However, some starburst dwarfs seem to have relatively symmetric
and unperturbed \hi disks; e.g. NGC~2915 \citep{Elson2011}
and VII~Zw~403 \citep{Simpson2011}. The relative fraction
of starburst dwarfs with symmetric/asymmetric \hi morphologies
in their outer regions is unclear, as well as the relation
between the extended \hi emission and the starburst activity.

\begin{table*}
\begin{minipage}{\textwidth}
\caption{Properties of the starburst.}
\small
\begin{center}
\begin{tabular}{l c c c c c c c c c}
\hline
Galaxy   & $b$ & SFR$_{0}$ & SFR$_{\rm{p}}$ & $\Sigma_{\rm SFR}(0)$ & $\Sigma_{\rm SFR}(t_{\rm p})$ & $\log(\rm{sSFR}_{0})$ & $\log(\rm {sSFR_{p}})$ & $t_{\rm{p}}$ & Ref. \\
         & & \multicolumn{2}{c}{(10$^{-3}$M$_{\odot}$ yr$^{-1}$)} & \multicolumn{2}{c}{(10$^{-3}$M$_{\odot}$ yr$^{-1}$ kpc$^{-2}$)} & \multicolumn{2}{c}{------ (Gyr$^{-1}$) ------} & (Myr) &  \\
\hline
NGC 625  & 3.0$\pm$0.1 &  4$\pm$2   & 86$\pm$20  & 0.12$\pm$0.06& 2.5$\pm$0.6 & -2.66$\pm$0.63 & -1.33$\pm$0.45 & 820$\pm$180& a \\
NGC 1569 & 21$\pm$1    & 80$\pm$15  & 240$\pm$10 & 2.8$\pm$0.5  & 8.5$\pm$0.3 & -1.79$\pm$0.21 & -1.31$\pm$0.11 & 40$\pm$10  & a \\
NGC 1705 & $\sim$6     & 314$\pm$78 & 314$\pm$78 & 44$\pm$11    & 44$\pm$11   & -0.65$\pm$0.56 & -0.65$\pm$0.56 & 3.0$\pm$1.5& b \\
NGC 2366 & 5.6$\pm$0.4 & 43$\pm$9   & 160$\pm$10 & 0.7$\pm$0.1  & 2.6$\pm$0.2 & -1.63$\pm$0.24 & -1.06$\pm$0.13 & 450$\pm$50 & a \\
NGC 4068 & 4.7$\pm$0.3 & 31$\pm$7   & 42$\pm$3   & 3.0$\pm$0.7  & 4.5$\pm$0.3 & -1.70$\pm$0.26 & -1.56$\pm$0.15 & 360$\pm$40 & a \\
NGC 4163 & 2.9$\pm$0.6 & 5.2$\pm$1.6& 12$\pm$3   & 1.6$\pm$0.5  & 3.8$\pm$0.9 & -2.13$\pm$0.43 & -1.77$\pm$0.39 &450$\pm$50 & a \\
NGC 4214 & 3.1$\pm$0.9 & 64$\pm$13  & 130$\pm$40 & 4.2$\pm$0.8  & 8.5$\pm$2.6 & -1.49$\pm$0.32 & -1.18$\pm$0.40 & 450$\pm$50 & a \\
NGC 4449 & 6.0$\pm$0.5 & 970$\pm$70 & 970$\pm$70 & 28$\pm$2     & 28$\pm$2    & -1.18$\pm$0.18 & -1.18$\pm$0.18 & 5$\pm$3    & a \\
NGC 5253 & 9.0$\pm$0.9 & 162$\pm$13 & 400$\pm$40 & 12$\pm$0.9   & 29$\pm$3    & -1.82$\pm$0.16 & -1.43$\pm$0.17 & 450$\pm$50 & a \\
NGC 6789 & 3.8$\pm$1.3 & 3.0$\pm$1.3& 15$\pm$5   & 1.9$\pm$0.8  & 9.7$\pm$3.2 & -2.21$\pm$0.52 & -1.51$\pm$0.44 & 565$\pm$65 & a \\
UGC 4483 &  14$\pm$3   & 11$\pm$4   & 11$\pm$2   & 9.7$\pm$1.8  & 8.8$\pm$3.5 & -0.80$\pm$0.27 & -0.84$\pm$0.45 & 565$\pm$65 & a \\
UGC 6456 &  7.6$\pm$1.1& 23$\pm$3   & 23$\pm$3   & 5.1$\pm$0.7  & 5.1$\pm$0.7 & -1.18$\pm$0.42 & -1.18$\pm$0.42 &  16$\pm$8  & a \\
UGC 6541 &  $\sim$3    & 3.0$\pm$1.5& ...        & 1.2$\pm$0.6  & ...         & -1.27$\pm$0.71 & ...            & ...        & c \\
UGC 9128 & 6.3$\pm$1.4 & 0.7$\pm$0.4& 5$\pm$1    & 0.6$\pm$0.3  & 4.4$\pm$0.9 & -2.11$\pm$0.59 & -1.26$\pm$0.25 & 150$\pm$50 & a \\
UGCA 290 & $\sim$3     & 11$\pm$8   & 42$\pm$15  & 4.3$\pm$3.1  & 16$\pm$6    & -0.80$\pm$0.88 & -0.22$\pm$0.61 & 15$\pm$5   & d \\
I Zw 18  & $\sim$30    & 100$\pm$50 & 100$\pm$50 & 127$\pm$64  & 127$\pm$64  & -0.07$\pm$0.73 & -0.07$\pm$0.73 & 10$\pm$5   & e \\
I Zw 36  & $\sim$7     &  25$\pm$12 & ...        & 9.8$\pm$4.7 & ...         & -0.35$\pm$0.69 & ...            & ...        & f \\
SBS 1415+437& $\sim$12 & 40$\pm$7   & 150$\pm$10 & 2.2$\pm$0.4 & 8.3$\pm$0.5 & -1.47$\pm$0.25 & -0.90$\pm$0.19 & 450$\pm$50  & a \\
\hline
\end{tabular}
\end{center}

\medskip
\textbf{Notes.} For a detailed description of these quantities, we refer to
Sect.~\ref{sec:AvsSB}. For UGC~6541 and I~Zw~36, the recent SFH ($\lesssim
1$ Gyr) is not well constrained due to the limited photometric depth of the
CMDs, thus we have no robust estimate of $\Sigma_{{\rm SFR}} (t_{\rm p})$,
sSFR$_{\rm p}$, and $t_{\rm p}$. Another problematic case is I~Zw~18, where
we prefer to use the SFR derived from H$\alpha$ observations rather than the
value derived from fitting the CMD. 

\medskip
\textbf{References.} (a)~\citet{McQuinn2010a}; (b)~\citet{Annibali2003};
(c)~\citet{SchulteLadbeck2000}; (d)~\citet{Crone2002}; (e)~\citet{Annibali2013};
(f)~\citet{SchulteLadbeck2001}.
\label{tab:SFRprop}
\end{minipage}
\end{table*}
In this paper we investigate the hypothesis that starbursts in dwarf
galaxies are triggered by external mechanisms. To this aim, we study in
detail the \hi emission in the outer regions of 18 starburst dwarfs.
The properties of our galaxy sample have been described in LVF14 and are
summarized in Tables \ref{tab:sample} and \ref{tab:SFRprop}. In short,
we selected galaxies that satisfy two criteria: (i) they have been resolved
into single stars by \textit{Hubble Space Telescope} (HST) observations,
providing detailed star-formation histories (SFHs) from the modelling of
color-magnitude diagrams \citep[CMDs; e.g.][]{McQuinn2010a}; and
(ii) they have a birthrate parameter $b=\rm{SFR/\overline{SFR}}\gtrsim3$,
identifying a starburst \citep{Kennicutt1998}. As far as we are aware of,
25 dwarf galaxies observed with the HST satisfy these criteria; we
collected new and archival \hi data for 18 of them. Note that the
birthrate parameter of a galaxy is generally very difficult to measure.
According to \citet{Lee2009}, galaxies with $b \gtrsim 2.5-3$ have
H$\alpha$ equivalent widths larger than $\sim$100~$\mathring{\rm A}$, and
constitute only $\sim6\%$ of the population of star-forming dwarfs at
$z=0$ (but see \citealt{McQuinn2010a} regarding the limitations of H$\alpha$
observations to identify starburst dwarfs). Here we consider birthrate
parameters derived directly from the SFHs as $\rm{SFR_{p}}/\overline{SFR}$,
where SFR$_{\rm p}$ is the peak SFR over the past 1 Gyr and $\overline{\rm
SFR}$ is the average SFR over the past 6 Gyrs \citep[see][]{McQuinn2010a}.
Thus, the HST information allows us to unambiguously identify starburst
dwarfs, which generally constitute a small subset of star-forming dwarfs,
and also to investigate possible relations between the \hi emission in
the outer galaxy regions and the detailed starburst properties (starburst
timescale, starburst intensity, etc.).

\section{Data analysis}\label{sec:dataAnal}

For the 18 galaxies in our sample, we collected both new and archival
21~cm-line observations from the VLA, the \textit{Westerbork Synthesis
Radio Telescope} (WSRT), and the \textit{Australia Telescope Compact
Array}  (ATCA). Archival observations were available for 15 galaxies,
mostly from the WHISP \citep{Swaters2002a}, THINGS \citep{Walter2008},
LITTLE-THINGS \citep{Hunter2012}, and VLA-ANGST \citep{Ott2012} projects
(see Table \ref{tab:data}). New \hi observations were obtained for the
remaining 3 objects using the VLA and the WSRT. In LVF14, we described
the reduction of these observations and presented \hi data at relatively
high spatial resolutions (ranging from 5$''$ to 30$''$ depending on the
individual galaxy properties). Here we analyse \hi datacubes at lower
spatial resolutions, which are more sensitive to low-column-density
gas in the outer regions (see Table \ref{tab:data}). We use the
same dataset as in LVF14 except for two objects: NGC~4449 and UGC~4483.
For NGC~4449, the \hi datacube from THINGS \citep{Walter2008} covers
a relatively small region on the sky, thus here we use the total \hi
map and velocity field from \citet{Hunter1998}, which were obtained
from VLA D-array observations in a 3$\times$3 pointing mosaic (covering
$\sim$1$^{\circ}$). For UGC~4483, in \citet{Lelli2012b} we reduced and
analysed archival \hi data obtained with the B- and C-arrays of the VLA,
but here we use the datacube from \citet{Ott2012} that includes also
new D-array observations, probing low-column-density gas on larger
angular scales.

\begin{table*}
\begin{minipage}{\textwidth}
\caption{Properties of the \hi datacubes.}
\small
\setlength{\tabcolsep}{5pt}
\begin{center}
\begin{tabular}{l c c c c c c c c c}
\hline
Galaxy       & Telescope & Source & \multicolumn{2}{c}{Smoothed Beam} & Ch. Sep.    & Taper & Rms Noise & \multicolumn{2}{c}{$N_{\hi}(3\sigma)$} \\
             &           &        &(asec$\times$asec)& (pc$\times$pc) &(km~s$^{-1}$)&       &(mJy/beam) & (10$^{20}$ cm$^{-2}$)&($M_{\odot}$ pc$^{-2}$) \\
\hline
NGC 625      & VLA       & a      & 30.0$\times$30.0 & 567$\times$567 & 2.6         & Hann. & 2.6      & 1.1$\pm$0.5          & 0.9$\pm$0.4  \\
NGC 1569     & VLA       & b      & 20.0$\times$20.0 & 330$\times$330 & 2.6         & Hann. & 1.1      & 1.6$\pm$0.7          & 1.3$\pm$0.6  \\
NGC 1705     & ATCA      & c      & 20.0$\times$20.0 & 494$\times$494 & 3.5         & Hann. & 1.2      & 1.1$\pm$0.4          & 0.9$\pm$0.3  \\
NGC 2366     & VLA       & b      & 20.0$\times$20.0 & 310$\times$310 & 2.6         & Hann. & 1.3      & 1.5$\pm$0.6          & 1.2$\pm$0.5  \\
NGC 4068     & WSRT      & d      & 30.0$\times$30.0 & 625$\times$625 & 2.0         & Unif. & 4.4      & 1.4$\pm$0.5          & 1.2$\pm$0.4  \\
NGC 4163     & VLA       & b      & 20.0$\times$20.0 & 290$\times$290 & 1.3         & Hann. & 1.3      & 0.9$\pm$0.3          & 0.8$\pm$0.3  \\
NGC 4214     & VLA       & b      & 30.0$\times$30.0 & 393$\times$393 & 1.3         & Hann. & 3.0      & 1.2$\pm$0.6          & 1.0$\pm$0.5  \\
NGC 4449     & VLA       & e      & 62.0$\times$54.0 &1262$\times$1099& 5.2         & Hann. & 1.3      & $\sim$0.7            & $\sim$0.6    \\
NGC 5253     & ATCA      & f      & 40.0$\times$40.0 & 679$\times$679 & 9.0         & Unif. & 2.7      & 1.0$\pm$0.3          & 0.8$\pm$0.3  \\
NGC 6789     & WSRT      & a      & 20.0$\times$20.0 & 349$\times$349 & 2.0         & Unif. & 1.3      & 1.0$\pm$0.3          & 0.8$\pm$0.3  \\
UGC 4483     & VLA       & g      & 20.0$\times$20.0 & 310$\times$310 & 2.6         & Hann. & 1.6      & 1.1$\pm$0.3          & 0.9$\pm$0.3  \\
UGC 6456     & VLA       & a      & 20.0$\times$20.0 & 417$\times$417 & 2.6         & Hann. & 2.4      & 1.6$\pm$0.7          & 1.3$\pm$0.6  \\
UGC 6541     & VLA       & b      & 20.0$\times$20.0 & 408$\times$408 & 1.3         & Hann. & 1.8      & 1.4$\pm$0.4          & 1.2$\pm$0.3  \\
UGC 9128     & VLA       & b      & 20.0$\times$20.0 & 213$\times$213 & 2.6         & Hann. & 1.6      & 1.3$\pm$0.4          & 1.0$\pm$0.3  \\
UGCA 290     & VLA       & a      & 20.0$\times$20.0 & 650$\times$650 & 1.6         & Unif. & 2.0      & 1.1$\pm$0.3          & 0.9$\pm$0.3  \\
I Zw 18      & VLA       & h      & 20.0$\times$20.0 &1765$\times$1765& 1.3         & Hann. & 1.0      & 1.0$\pm$0.6          & 0.8$\pm$0.4  \\
I Zw 36      & VLA       & b      & 20.0$\times$20.0 & 572$\times$572 & 2.6         & Hann. & 1.9      & 1.3$\pm$0.4          & 1.0$\pm$0.3 \\
SBS 1415+437 & VLA       & a      & 20.0$\times$20.0 &1319$\times$1319& 1.6         & Unif. & 2.8      & 1.3$\pm$0.3          & 1.0$\pm$0.3 \\
\hline
\end{tabular}
\end{center}

\medskip
\textbf{References.} (a)~LVF14; (b)~\citet{Hunter2012}; (c)~\citet{Elson2013};
(d)~\citet{Swaters2002a}; (e)~\citet{Hunter1998}; (f)~\citet{LopezSanchez2012};
(g)~\citet{Ott2012}; (h)~\citet{Lelli2012a}.
\label{tab:data}
\end{minipage}
\end{table*}
For every galaxy, we chose the optimal spatial resolution using the
following approach. We first inspected the \hi cube at the highest
spatial and spectral resolutions available. Then, this cube was
progressively smoothed in the image plane to 10$''$, 20$''$, 30$''$,
and 40$''$, and total \hi maps at different spatial resolutions were
constructed by summing masked channel maps. The smoothing procedure
was halted when the total \hi map reached a 3$\sigma$ column density
sensitivity of $\sim$10$^{20}$ atoms~cm$^{-2}$, which is adequate
to investigate the \hi morphology in the outer regions \citep[e.g.]
[]{Swaters2002a} and, at the same time, allows us to preserve a
relatively high angular resolution (typically 20$''$ except for
4 cases, see Table~\ref{tab:data}). The masks were obtained by
smoothing the cubes in velocity to $\sim$10 km~s$^{-1}$ and in
the image plane to 1$'$ (2$'$ for NGC~2366, NGC~4214, and NGC~5253
given their large angular extent), and clipping at 3$\sigma_{1'}$
($\sigma_{1'}$ is the rms noise in the smoothed cube). For NGC~4214,
the cube was smoothed in velocity to only $\sim$2.6 km~s$^{-1}$
because only a few line-free channels were available at its
high-velocity end. All the masks were visually inspected; residual
noise peaks and Galactic emission were interactively blotted out.
Note that the original, high-resolution cubes were obtained using
a robust weighting technique \citep{Briggs1995} with robust parameter
$\Re \simeq 0$, thus they have relatively low column density
sensitivity but their beam profile is close to a Gaussian shape.
We avoided using natural-weighted datacubes because the broad
wings of their beam profiles may lead to spurious detections of
diffuse emission, especially when the \hi data are not cleaned
down to the noise level (as is the case for the LITTLE-THINGS
datacubes that are cleaned down to only 2.5$\sigma$, see
\citealt{Hunter2012}).

Since we are interested in low-column-density \hi emission, it
is important to accurately estimate the 3$\sigma$ column density
sensitivity of the total \hi maps. The noise in a total \hi map is
not uniform but varies from pixel to pixel because at each spatial
position one adds a different number of channels, given that only
the pixels inside a given mask are considered. Following \citet[]
[]{Verheijen2001}, we constructed signal-to-noise maps and calculated
a pseudo-3$\sigma$ column density contour $N_{\hi}(3\sigma)$ by
averaging the values of the pixels with signal-to-noise ratio
between 2.75 and 3.25. We also calculated the rms around the
mean value of these pixels to estimate the uncertainty on
$N_{\hi}(3\sigma)$. In particular, we halted the progressive
smoothing of the \hi cubes when the value of $N(3\sigma)$ was
equal to $1\times10^{20}$ cm$^{-2}$ within the errors. The
derivation of the signal-to-noise maps is described in detail
in Appendix~\ref{app:noise}.

We calculated total \hi fluxes from the smoothed maps by considering
the pixels with a flux density higher than $\frac{1}{2} N_{\hi}
(3\sigma)$, that can be considered as a pseudo-1.5$\sigma$ contour
(see Table~\ref{tab:HIprop}). Our \hi fluxes are in overall agreement
with those from single-dish observations: the differences are typically
$\lesssim$15$\%$ apart for two objects (NGC~1569 and UGC~6456) that
are affected by Galactic emission. Our smoothed \hi maps, thus,
recover most of the \hi emission from the galaxy.

We also derived \hi velocity fields by estimating an intensity-weighted
mean (IWM) velocity from the masked datacube at the optimal resolution,
clipping at 2$\sigma$ and considering only the pixels within the
pseudo-3$\sigma$ contour of the total \hi map. Since the \hi profiles
are generally broad and asymmetric, these low-resolution IWM velocity
fields are uncertain and provide only an overall description of kinematics
of the extended gas. For the 18 galaxies in our sample, a detailed analysis
of the \hi kinematics has been presented in \citet{Lelli2012a, Lelli2012b}
and LVF14.

\begin{table*}
\begin{minipage}{\textwidth}
\caption{\hi properties.}
\small
\begin{center}
\begin{tabular}{l c c c c c c c}
\hline
Galaxy       & $S_{\hi}$& $M_{\hi}$ &\multicolumn{2}{c}{$E_{\hi}$} & $E_{\hi}/R_{\rm{opt}}$ & $t_{E_{\hi}}$ & $A$\\
  & (Jy km/s)  & ($10^{7} M_{\odot}$) &(amin)&(kpc)&     &(Gyr) &      \\
\hline
NGC 625      & 27.0     & 9.7$\pm$2.2 &  4.7 & 5.4 & 1.6 & 1.1  & 0.60 \\
NGC 1569     & 106.6    & 29.1$\pm$4.5& 10.8 &10.7 & 3.6 & 1.3  & 0.68 \\
NGC 1705     & 18.2     & 11.1$\pm$2.9& 4.0  & 5.9 & 3.9 & 0.5  & 0.74 \\
NGC 2366     & 254.8    & 62$\pm$17   &10.1  & 9.4 & 2.1 & 1.1  & 0.53 \\
NGC 4068     & 34.2     & 14.9$\pm$1.6& 3.6  & 4.5 & 2.5 & 0.7  & 0.52 \\
NGC 4163     & 7.2      & 1.5$\pm$0.2 & 2.3  & 2.0 & 2.0 & 1.2  & 0.64 \\
NGC 4214     & 250.8    & 43$\pm$8    & 10.4 & 8.2 & 3.7 & 0.6  & 0.52 \\
NGC 4449     &721.2&300$\pm$77  & 34.3 & 42  & 13  & 7.4  & 0.77 \\
NGC 5253     & 47.6     & 13.8$\pm$3.4& 4.6  & 4.7 & 2.2 & ...  & 0.57 \\
NGC 6789     & 5.9      & 1.8$\pm$0.3 & 1.8  & 1.9 & 2.7 & 0.2  & 0.60 \\
UGC 4483     & 12.0     & 2.9$\pm$0.5 & 2.7  & 2.5 & 4.2 & 0.8  & 0.73 \\
UGC 6456     & 10.4     & 4.5$\pm$0.5 & 1.9  & 2.4 & 1.9 & 1.5  & 0.51 \\
UGC 6541     & 2.8      & 1.2$\pm$0.2 & 1.9  & 2.3 & 2.5 & ...  & 0.73 \\
UGC 9128     & 11.1     & 1.3$\pm$0.2 & 1.8  & 1.2 & 2.0 & 0.3  & 0.48 \\
UGCA 290     & 1.35     & 1.4$\pm$0.2 & 1.1  & 2.1 & 2.3 & ...  & 0.70 \\
I Zw 18      & 2.7      & 21$\pm$4    & 1.6  & 8.5 & 17  & 1.4  & 0.76 \\
I Zw 36      & 8.2      & 6.7$\pm$1.3 & 1.9  & 3.3 & 3.7 & 0.7  & 0.64 \\
SBS 1415+437 & 4.6      & 20.1$\pm$4.6& 1.8  & 7.3 & 3.0 & 2.5  & 0.42 \\
\hline
\end{tabular}
\end{center}

\medskip
\textbf{Notes.} The VLA data of NGC~4449 seem to miss diffuse \hi emission
\citep{Hunter1998}, thus $S_{\hi}$ and $M_{\hi}$ may be underestimated.
The asymmetry parameter $A$ is calculated using the total \hi maps in
Fig.~\ref{fig:mosaic1} (see Tab.~\ref{tab:data} for the relative spatial
resolutions and column density sensitivities).
\label{tab:HIprop}
\end{minipage}
\end{table*}
\section{The \hi gas in the outer regions}\label{sec:morpho}

In the following we qualitatively describe the overall properties
of the \hi gas in the outer regions of starburst dwarfs, while
in Sect.~\ref{sec:asym} we quantify the outer \hi asymmetry
using a new asymmetry parameter; we also make a comparison with
a control sample of typical star-forming Irrs and investigate
the relation between outer \hi distribution and starburst
properties. In Sect.~\ref{sec:indi} we then describe each
individual galaxy in detail and discuss its nearby environment.

\begin{figure*}
\centering
\includegraphics[width=0.9\textwidth]{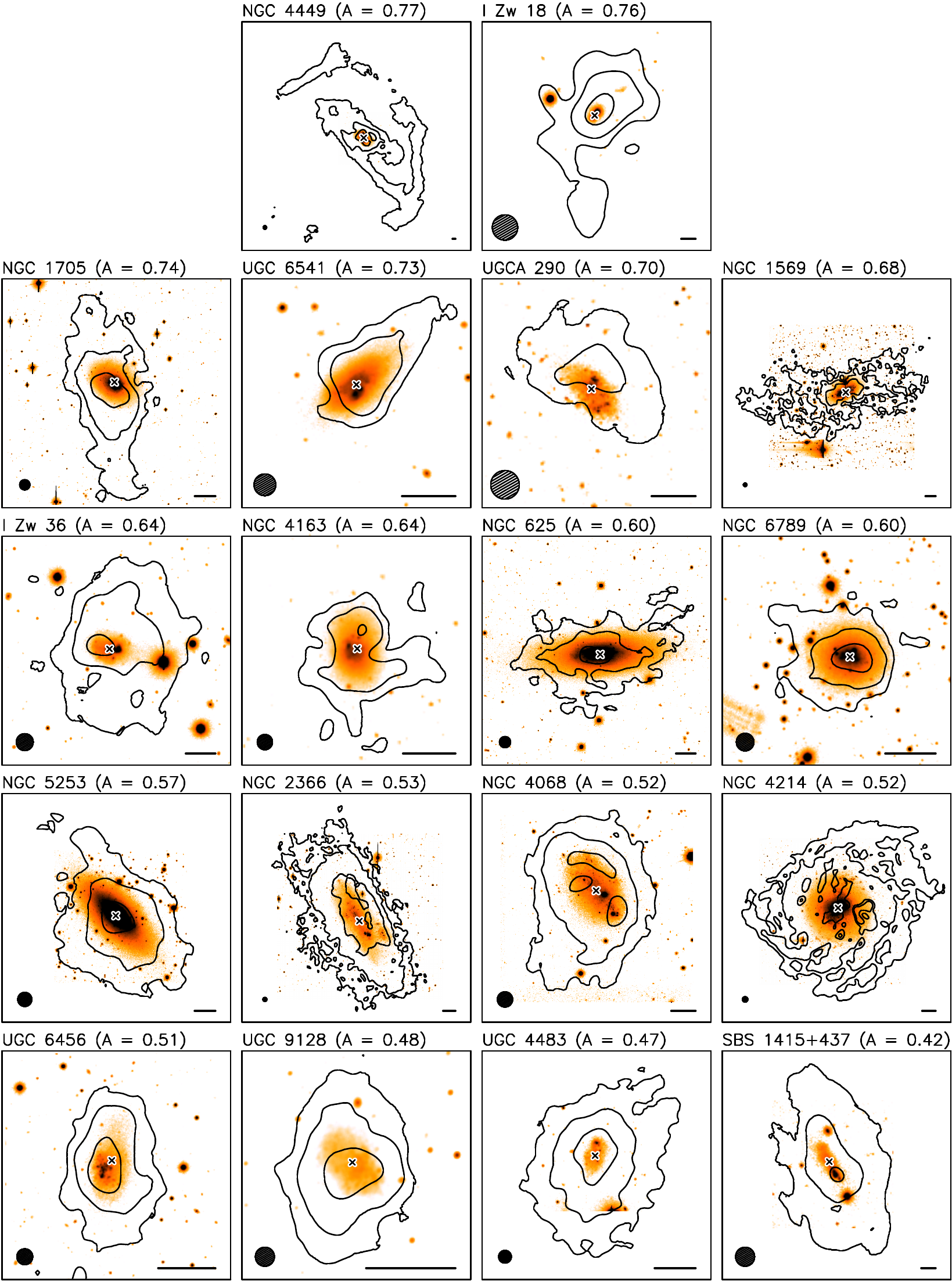}
\caption{Total \hi maps superimposed on optical images. The galaxies
are ordered according to the value of the asymmetry parameter $A$
(see Sect.~\ref{sec:asym} for details). The contours are at 1, 4,
16$\times$10$^{20}$ atoms~cm$^{-2}$. The cross shows the optical
center. The ellipse to the bottom-left shows the \hi beam
(given in Table~\ref{tab:data}). The bar to the bottom-right
corresponds to 1~kpc.}
\label{fig:mosaic1}
\end{figure*}

Figure~\ref{fig:mosaic1} shows the total \hi maps of our 18
starburst dwarfs superimposed on optical images; in each map
the \hi contours correspond to 1, 4, and 16$\times$10$^{20}$
atoms~cm$^{-2}$. The outer \hi distribution of starburst dwarfs
shows a variety of morphologies. Several galaxies have heavily
disturbed \hi distributions, characterized by large-scale
asymmetries, long filaments, and/or a large optical-\hi offset
(NGC~4449,  I~Zw~18, NGC~1705, UGC~6541, NGC~1569, and I~Zw~36).
Other galaxies, instead, show lopsided \hi morphologies, characterized
by minor asymmetries and/or extensions in the outer parts
(NGC~2366, NGC~4068, NGC~4214, UGC~6456, UGC~9128, UGC~4483,
and SBS~1415+437). There is \textit{not} a clear-cut separation
between these two types of \hi morphologies, since there are
several ``intermediate'' cases that have relatively regular
\hi distributions in the inner parts and small tails/filaments
in the outer regions (NGC~4163, NGC~625, NGC~6789, and NGC~5253).

\begin{figure*}
\centering
\includegraphics[width=0.9\textwidth]{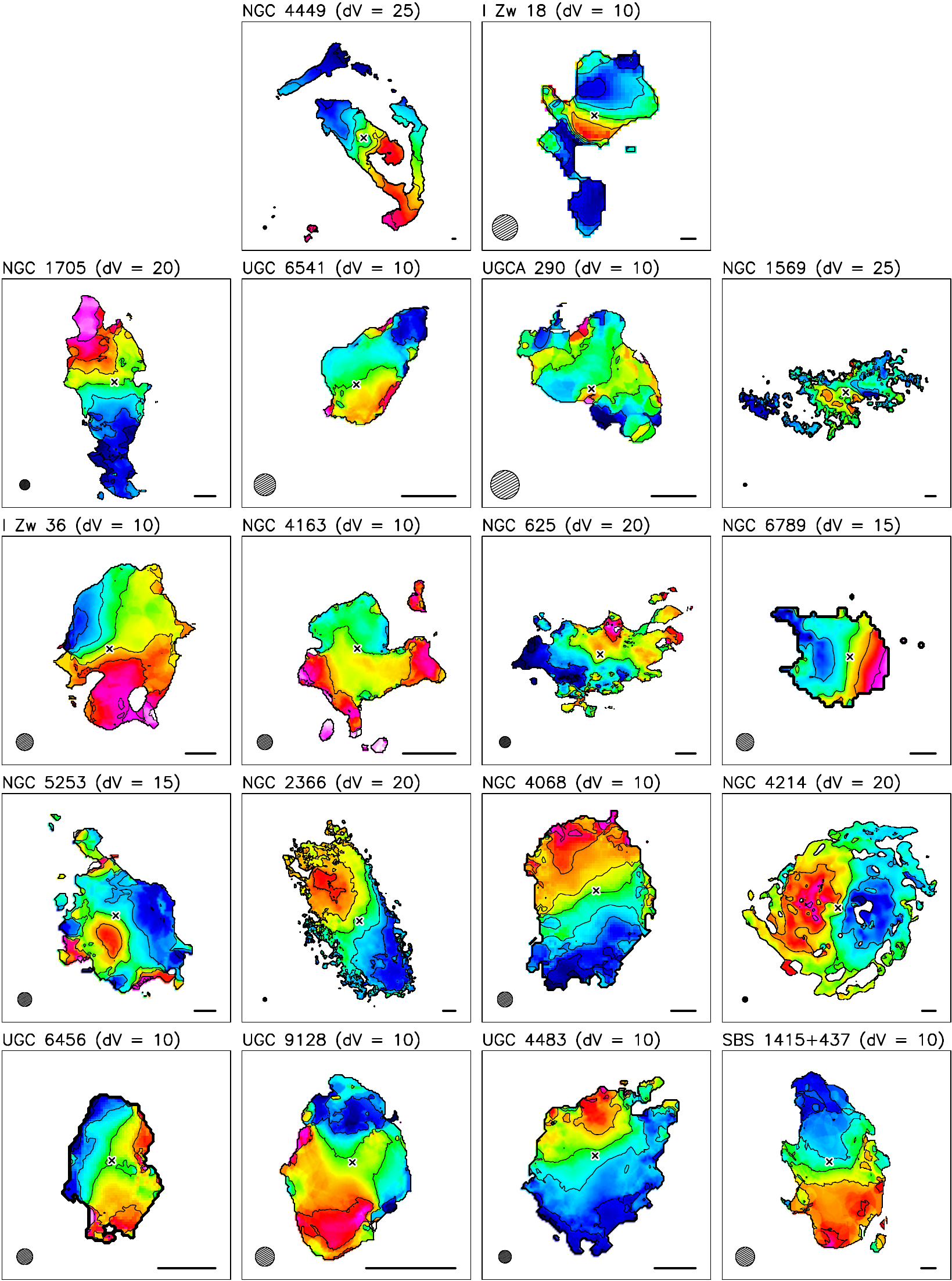}
\caption{\hi velocity fields. The box sizes are the same as in
Fig.~\ref{fig:mosaic1}. The velocity separation $dV$ between
the contours is indicated. The cross shows the optical center.
The ellipse to the bottom-left shows the \hi beam (given in
Table~\ref{tab:data}). The bar to the bottom-right corresponds
to 1~kpc.}
\label{fig:VF}
\end{figure*}

We estimated the extent of the \hi distribution $E_{\hi}$ by
measuring the projected distance between the optical center
of the galaxy and the outermost pixel with an observed column
density of 1$\times$10$^{20}$ atoms~cm$^{-2}$. Note that $E_{\hi}$
is conceptually different from the \hi radius $R_{\hi}$, as the
latter is defined as the radius where the azimuthally averaged
\hi surface density profile (corrected for inclination) reaches
1~M$_{\odot}$~pc$^{-2}$ ($\sim$1.2$\times$10$^{20}$ atoms~cm$^{-2}$;
see e.g. \citealt{Swaters2002a}). Since $E_{\hi}$ is \textit{not}
obtained from an azimuthal average over the total \hi map,
it properly takes into account anomalous extensions in the
\hi distribution (such as tails or filaments), but it may
be affected by projection effects along the line of sight.
Table~\ref{tab:HIprop} lists the values of $E_{\hi}$ and
$E_{\hi}/R_{\rm opt}$, where the optical radius $R_{\rm opt}$
is defined as 3.2 exponential scale-lengths (see LVF14)
and is given in Table~\ref{tab:sample}. For the 18 galaxies
in our sample, $E_{\hi}/R_{\rm opt}$ ranges from $\sim$1.5
to $\sim$4, except for I~Zw~18 ($E_{\hi}/R_{\rm opt} \simeq 17$)
and NGC~4449 ($E_{\hi}/R_{\rm opt}\simeq13$). These two objects
show exceedingly extended \hi tails with relatively high column
densities ($\sim1-2\times$10$^{20}$ atoms~cm$^{-2}$). Intriguingly,
both I~Zw~18 and NGC~4449 have a companion galaxy with $L_{\rm R}
\simeq 0.5 - 1 \times 10^{7}$~$L_{\odot}$ at a projected distance
$\lesssim10$~kpc. For I~Zw~18, there are strong indications that
the extended \hi emission is associated with the secondary body
\citep[see][]{Lelli2012a}. For NGC~4449, instead, the relation
between the companion galaxy and the long \hi filaments is
unclear \citep[see][]{Delgado2012}.

Finally, we describe the \hi kinematics in the outer regions.
Figure~\ref{fig:VF} shows the velocity fields of our 18 galaxies.
As we stressed in Sect.~\ref{sec:dataAnal}, these velocity
fields are only indicative due to the complex structure of
the \hi profiles, but they provide an overall description of
the gas kinematics. For starburst dwarfs with a rotating \hi disk,
the outer gas generally is kinematically connected to the inner
\hi distribution (except for I~Zw~18, discussed in \citealt{Lelli2012a}).
This suggests that the outer disks can be regularized in a
few orbital times by differential rotation. We calculated
the orbital times $t_{\rm E_{\hi}}$ at $E_{\hi}$ using the
rotation velocities in Table~\ref{tab:sample} (from LVF14).
These rotation velocities are typically estimated at $\sim$1
to 2 $R_{\rm opt}$, thus we are extrapolating their values to
larger radii by assuming that the rotation curve is flat and
the outer gas lies approximately in the same plane as the inner
\hi disk. The values of $t_{\rm E_{\hi}}$ in Table~\ref{tab:HIprop},
therefore, should be considered as \textit{order-of-magnitude}
estimates. Despite these uncertainties, the orbital times at
$E_{\hi}$ are consistently of the order of $\sim$0.5 to $\sim$1 Gyr
(except for NGC~4449 with $t_{\rm E_{\hi}}\simeq 7$~Gyr), indicating
that the outer asymmetries must be relatively recent and possibly
short-lived.

\section{Quantifying the \hi asymmetry}\label{sec:asym}

\subsection{The asymmetry parameter}\label{sec:Aout}

To investigate the relation between the \hi morphology and
the starburst, it is desirable to quantify the degree of
asymmetry/lopsidedness in the outer \hi distribution of each
individual galaxy. The infrared/optical morphologies of
galaxies are usually quantified using the Concentration-Asymmetry-Smoothness
(CAS) parameters \citep[][]{Conselice2003} and the Gini-$M_{20}$
parameters \citep[][]{Lotz2004}. Recently, \citet{Holwerda2011a,
Holwerda2011b, Holwerda2013} used these parameters to quantify
the \hi morphologies in several samples of nearby galaxies. In
particular, \citet{Holwerda2011b} used total \hi maps from the
WHISP survey, and found that the CAS and Gini parameters only
weakly correlate with previous \textit{visual} classifications
of morphological lopsidedness (by \citealt{Swaters2002a} and
\citealt{Noordermeer2005}). \citet{Holwerda2011a, Holwerda2011b}
defined the asymmetry parameter $\cal A$ as:
\begin{equation}
{\cal A} = \dfrac{\sum_{i, \, j}|I(i,j) - I_{180^{\circ}} (i, j)|}{\sum_{i, j}|I(i,j)|},
\end{equation}
where $I(i,j)$ and $I_{180}(i, j)$ are the flux densities at
position $(i, j)$ in the original image and in an image rotated
by 180$^{\circ}$ with respect to the galaxy center, respectively.
This definition normalizes the residuals between the original
image and the rotated image to the \textit{total} flux. Thus,
asymmetries in the outer regions may have negligible weight
in the sum, since the flux densities in the outer parts can
be up to $\sim$2 orders of magnitude lower than those in
the inner parts. This effect has been pointed out by
\citet{Holwerda2013} (see their Sect.~3.2 and Fig.~2),
who demonstrated that the value of $\cal A$ does not strongly
depend on the outer \hi emission observed in low-resolution
\hi maps. Our goal, instead, is to give weight to the
large-scale asymmetries in the outer parts. Thus, we use
a new definition of $A$:
\begin{equation}\label{eq:Aout}
 A = \dfrac{1}{N} \sum^{N}_{i, j} \dfrac{|I(i,j) - I_{180^{\circ}}(i,j)|}{|I(i,j)+I_{180^{\circ}}(i,j)|},
\end{equation}
where $N$ is the total number of pixels in the image. This definition
normalizes the residuals at position $(i, j)$ to the \textit{local}
flux density. In particular, if \hi emission is detected only on one
side of the galaxy, the residuals at $(i,j)$ and $(i,j)_{180^{\circ}}$
get the maximum value ($= 1$).

In Fig.~\ref{fig:mosaic1} the total \hi maps of our 18 starburst
dwarfs are ordered according to the value of $A$. It is clear that
our definition of $A$ reliably quantifies the \hi asymmetry in
the outer parts. The value of $A$, however, may depend on (i)
the assumed galaxy center, (ii) the column density sensitivity
of the \hi observations, and (iii) the spatial resolution in
terms of both the beam-size in kpc and the relative number of
beams across the \hi map. In the following, we describe the
effects of these factors on the value of $A$.

We adopted the optical centers derived in LVF14 by fitting ellipses
to the outer isophotes (see Table~\ref{tab:sample}). We did not
consider the \textit{kinematic} centers because 50$\%$ of the
galaxies in our sample have either a kinematically-disturbed
\hi disk or an unsettled \hi distribution, thus the kinematic
parameters are either very uncertain or undefined. Moreover,
the use of the optical center returns high values of $A$ for
galaxies that show a strong offset between the \hi distribution
and the stellar body (e.g. NGC~1705 and UGCA~290), that may
indicate a recent interaction/accretion event. We checked
that small changes in the value of the centers ($\sim$2$''$)
do not significantly affect the value of $A$ (the absolute
differences in $A$ are $\lesssim0.05$). This is expected because
(i) the typical uncertainties on the position of the optical center
($\sim1''$ to 2$''$) are much smaller than the \hi beam ($\gtrsim
20''$), and (ii) by our definition of $A$, high-column-density
asymmetries in the inner parts do not have much weight.

\begin{figure}
\centering
\includegraphics[width=0.45\textwidth]{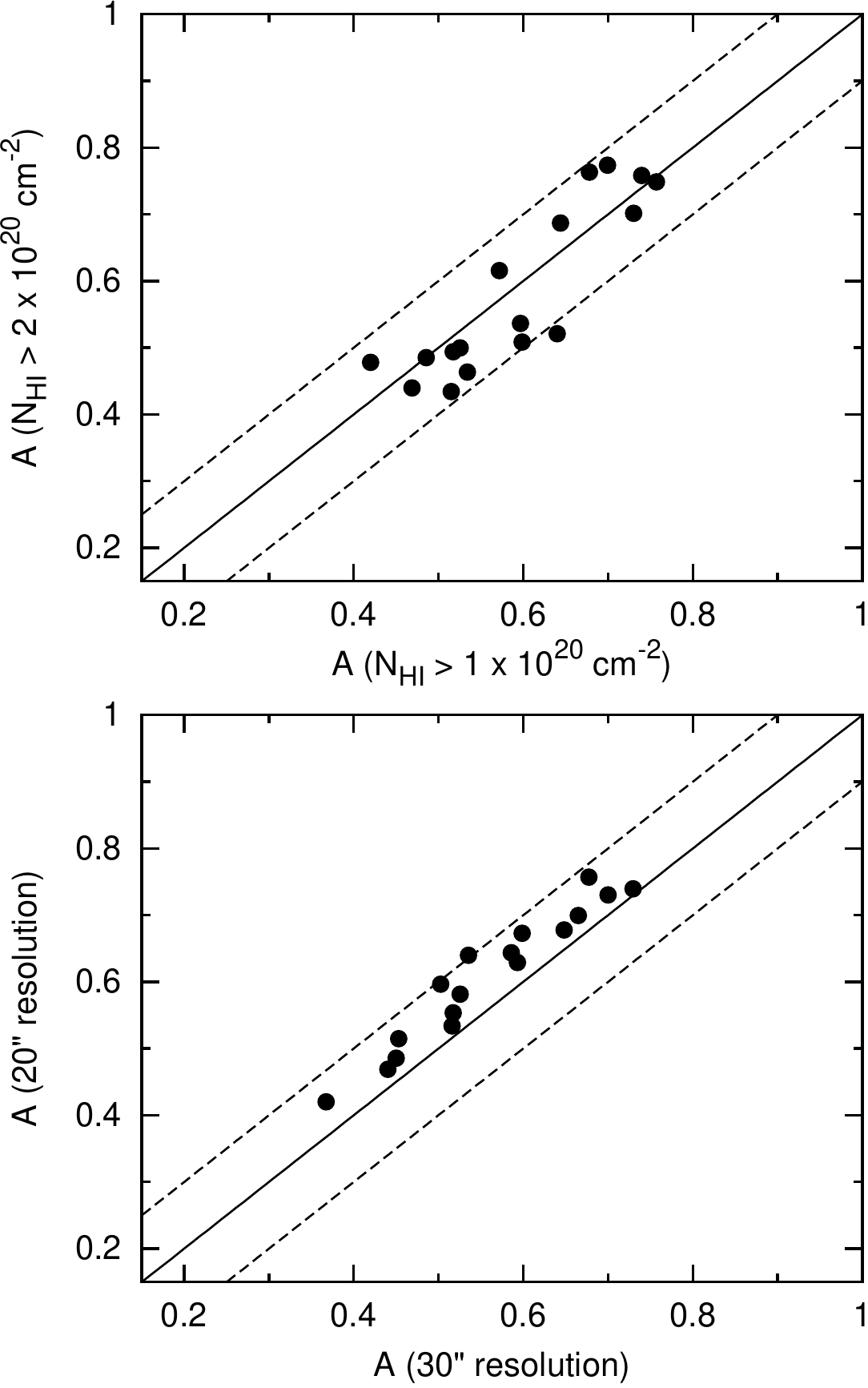}
\caption{Testing the dependence of the asymmetry parameter $A$ on the
column-density threshold (\textit{top}) and on the spatial resolution
of the total \hi maps (\textit{bottom}). In the top panel, dots show
the values of $A$ obtained from the total \hi maps in Fig.~\ref{fig:mosaic1}
using a column-density threshold of 1 and $2\times10^{20}$ atoms cm$^{-2}$.
In the bottom panel, dots show the values of $A$ using \hi maps at
20$''$ and 30$''$ (keeping the column-density threshold fixed to
10$^{20}$ atoms~cm$^{-2}$). The solid line corresponds to a null
variation in the value of $A$, whereas the dashed lines correspond
to variations  of $\pm$0.1. See Sect.~\ref{sec:Aout} for details.}
\label{fig:Atest}
\end{figure}
Regarding the column density sensitivity, the values of $A$ in 
Fig.~\ref{fig:mosaic1} and Table~\ref{tab:HIprop} have been
calculated considering only pixels with $N_{\hi} \geq 10^{20}$~cm$^{-2}$,
since this corresponds to the pseudo-3$\sigma$ contour of our total
\hi maps (see Sec.~\ref{sec:dataAnal} and Appendix~\ref{app:noise}).
To test the effect of this column density threshold, in Fig.~\ref{fig:Atest}
(top) we compare the values of $A$ obtained by considering thresholds
of 1 and 2$\times$10$^{20}$ cm$^{-2}$. The differences in $A$ are
within $\sim$0.1 and do not show a systematic trend, implying that
the detailed shape of the outermost contour in the total \hi map
does not strongly affect the value of $A$. We warn, however, that
the use of a column-density threshold provides reliable results
as long as (i) one does \textit{not} consider values much below the
pseudo-3$\sigma$ contour, introducing noise in the total \hi maps,
and (ii) one does \textit{not} consider high column density thresholds
(e.g. $N_{\hi}\gtrsim 5-10 \times 10^{20}$ cm$^{-2}$), probing the
small-scale clumpiness of the \hi distribution. We also note that,
when using a fixed column-density threshold for different galaxies, the
inclination $i$ of the \hi disk may introduce some systematic effects,
given that \textit{projected} column densities correspond to different
\textit{face-on} surface densities. For an optically-thin \hi disk,
the projected column density increases with $1/\cos(i)$. Thus,
inclination effects on the column-density threshold become important
only in edge-on disks with $i \gtrsim 70^{\circ}$, for which projected
column densities of $\sim 1 \times 10^{20}$~cm$^{-2}$ would correspond
to face-on surface densities that are lower by a factor $\gtrsim 3$.
In our galaxy sample, the inclinations of the \hi disks are $\lesssim
70^{\circ}$ (see LVF14), thus projection effects are not a serious
concern here.

The spatial resolution of the \hi observations deserves some attention
in the derivation of $A$. To quantify the effects of beam smearing,
we constructed total \hi maps at 20$''$ and 30$''$ resolution for
all the galaxies in our sample (except for NGC~4449 that has \hi
data at a native resolution of $\sim$60$''$). The respective values
of $A$, calculated using a threshold of 1$\times$10$^{20}$ cm$^{-2}$,
are compared in Fig.~\ref{fig:Atest} (bottom panel). As expected,
total \hi maps at higher resolutions systematically yield higher
values of $A$. The differences in $A$, however, appear reasonably
small (within 0.1). Galaxies with a small number of beams along
the major axis of the \hi disk (e.g. NGC~4163, NGC~6789, I~Zw~18)
typically show the largest differences in $A$ ($\sim$0.1),
whereas galaxies with well-resolved \hi maps (e.g. NGC~2366,
NGC~1569, and NGC~4214) show very small differences ($\lesssim$0.03).
For the latter galaxies, a severe smoothing of the \hi data down
to 60$''$ (a factor 3) would still give differences in $A$
$\lesssim0.1$. Thus, we draw the following conclusions: (i)
to have a reliable estimate of $A$, one needs at least $\sim$5
resolution elements along the major axis of the total \hi map,
and (ii) when the previous condition is met, differences in
spatial resolution by a factor of $\sim$3 give relatively small
differences in $A$ ($\lesssim0.1$). Our total \hi maps are all
reasonably resolved (see Fig.~\ref{fig:mosaic1}) and have
linear resolutions ranging from $\sim$0.3 to $\sim$0.7 kpc
(see Table~\ref{tab:data}), thus it makes sense to compare
the values of $A$ for different galaxies. Exceptions are NGC~4449,
I~Zw~18, and SBS~1415+437, that have total \hi maps with linear
resolutions $\gtrsim$1~kpc. Despite the low linear resolution,
NGC~4449 and I~Zw~18 show the highest values of $A$ in our
sample, indicating that data at higher resolutions would only
increase the difference with the other galaxies. On the contrary,
SBS~1415+437 has the lowest value of $A$ in our sample; this
may be an effect of beam-smearing. We did not build total
\hi maps at the same linear resolution (in kpc) for all
the galaxies because it is not possible to find a compromise
between the required number of beams along the \hi major
axis ($\gtrsim5$ in order to have a proper estimate of $A$)
and the 3$\sigma$ column density sensitivity ($\lesssim 1
\times 10^{20}$ in order to probe the outer \hi emission).

\subsection{Comparison with typical irregulars}\label{sec:AoutComp}

In this section, we estimate $A$ for a control sample of typical
star-forming Irrs and make a comparison with our sample of starburst
dwarfs. We use total \hi maps from the VLA-ANGST survey \citep{Ott2012},
which provides multi-configuration VLA observations for 29 low-mass
galaxies from the \textit{Advanced Camera for Surveys Nearby Galaxy
Survey Treasury} (ANGST; \citealt{Dalcanton2009}). In order to have
two galaxy samples that span similar ranges of stellar and \hi masses,
we exclude 8 objects with $M_{\rm B}\lesssim-11$ (nearly equivalent
to $M_{*}\lesssim 10^{7}$~M$_{\odot}$), given that such very low-mass
galaxies are not present in our sample of starburst dwarfs. We also
exclude AO~0952+69 (Arp's loop), that may be a feature in the spiral
arm of M81 \citep{Ott2012} or a tidal dwarf galaxy \citep{Weisz2011}.
The VLA-ANGST sample also contains 3 starburst dwarfs that are included
in our sample (NGC~4163, UGC~4483, and UGC~9128), which we use to test
the consistency between our total \hi maps and the VLA-ANGST ones.
The control sample of typical Irrs, therefore, contains 17 galaxies.

The starburst and control samples cover similarly broad ranges
of absolute magnitudes ($-12 \lesssim M_{\rm B} \lesssim -18$)
and \hi masses ($10^{7} \lesssim M_{\hi}/M_{\odot} \lesssim 10^{9}$).
For the starburst sample, the mean values in $M_{\rm B}$ and $M_{\hi}$
are, respectively, $-14.9$ mag and $3.0 \times 10^{8}$~$M_{\odot}$,
while for the control sample they are $-13.9$~mag and $1.1 \times 10^{8}$
$M_{\odot}$. Despite the starburst dwarfs are, on average, slightly
more luminous and gas-rich than the typical Irrs, it is clear that
the two samples can be properly compared. Moreover, similarly to
the starburst dwarfs in our sample, the VLA-ANGST galaxies have
been resolved into single stars by HST \citep{Dalcanton2009}.
For most of these galaxies, \citet{Weisz2011} derived SFHs by
averaging the SFR over a single time-bin in the last $\sim$1~Gyr,
thus we cannot check whether they have a recent birthrate
parameter $b \lesssim 3$, confirming that they are \textit{not}
starburst dwarfs. However, as far as we are aware of, the 17
galaxies in our control sample do not show any sign of recent
starburst activity either in their CMDs nor in their integrated
spectra, thus we consider them representative for typical
star-forming Irrs.

\begin{figure}
\centering
\includegraphics[width=0.45\textwidth]{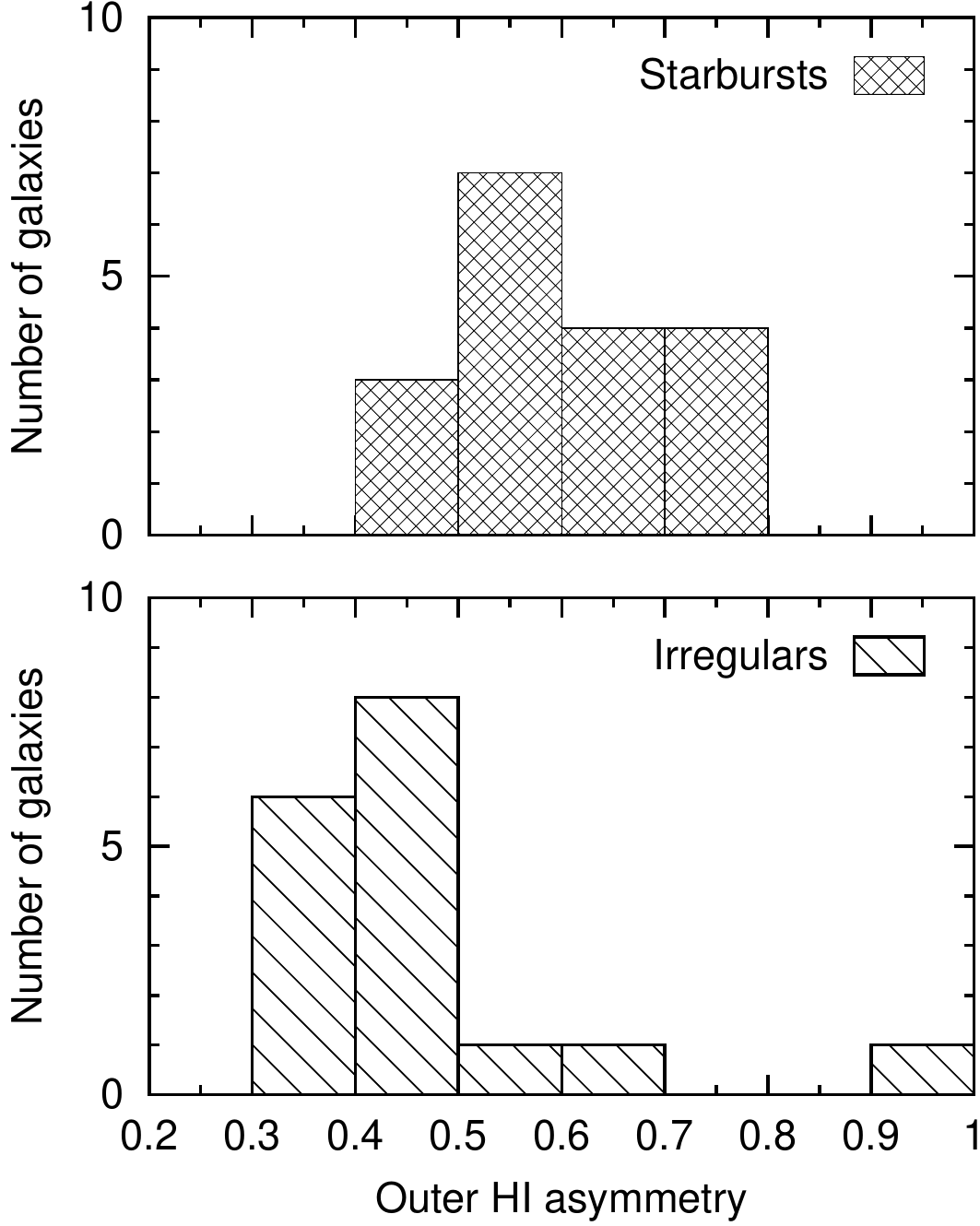}
\caption{The asymmetry parameter $A$ for our sample of 18 starburst
dwarfs and for a control sample of 17 typical Irrs, drawn from the
VLA-ANGST survey. Starburst dwarfs clearly have more asymmetric \hi
morphologies in the outer parts than typical Irrs of similar masses.
See Sect.~\ref{sec:AoutComp} for details.}
\label{fig:Ahisto}
\end{figure}
The natural-weighted \hi maps from VLA-ANGST have both an adequate
number of resolution elements along the \hi major axis and a 3$\sigma$
column density sensitivity $\lesssim10^{20}$ cm$^{-2}$ \citep[see][]
{Ott2012}, thus we can safely calculate $A$ using a column-density
threshold of 10$^{20}$ cm$^{-2}$ (as for our starburst dwarfs).
The beam-sizes of these \hi maps range from $\sim$60 to $\sim$200~pc,
significantly smaller than those of our own \hi maps (cf. with
Table~\ref{tab:data}). As we discussed in Sect.~\ref{sec:Aout},
this implies that the values of $A$ for our starburst dwarfs may
be systematically \textit{underestimated} with respect to those of
the VLA-ANGST Irrs. For UGC~4483 and UGC~9128, however, the VLA-ANGST
\hi maps and our total \hi maps yield remarkably consistent results:
UGC~4483 has $A=0.469$ from our map and $A=0.464$ from the VLA-ANGST
one, while UGC~9128 has $A=0.485$  from our map and $A=0.480$ from
the VLA-ANGST one. NGC~4163, instead, shows a significant discrepancy:
our map yields $A=0.64$ while the VLA-ANGST one returns $A=0.50$.
The VLA-ANGST map of NGC~4163 does not trace the full extent of
the \hi tail to the West and does not detect the cloud complexes
to the South (compare our Fig.~\ref{fig:mosaic1} with Fig.~18
of \citealt{Ott2012}). This is likely due to a different masking
of the \hi emission during the derivation of the total \hi map.
We are confident that these \hi features are real given that they
have been detected also by \citet{Swaters2002a} using WSRT data.

\begin{figure*}
\centering
\includegraphics[width=\textwidth]{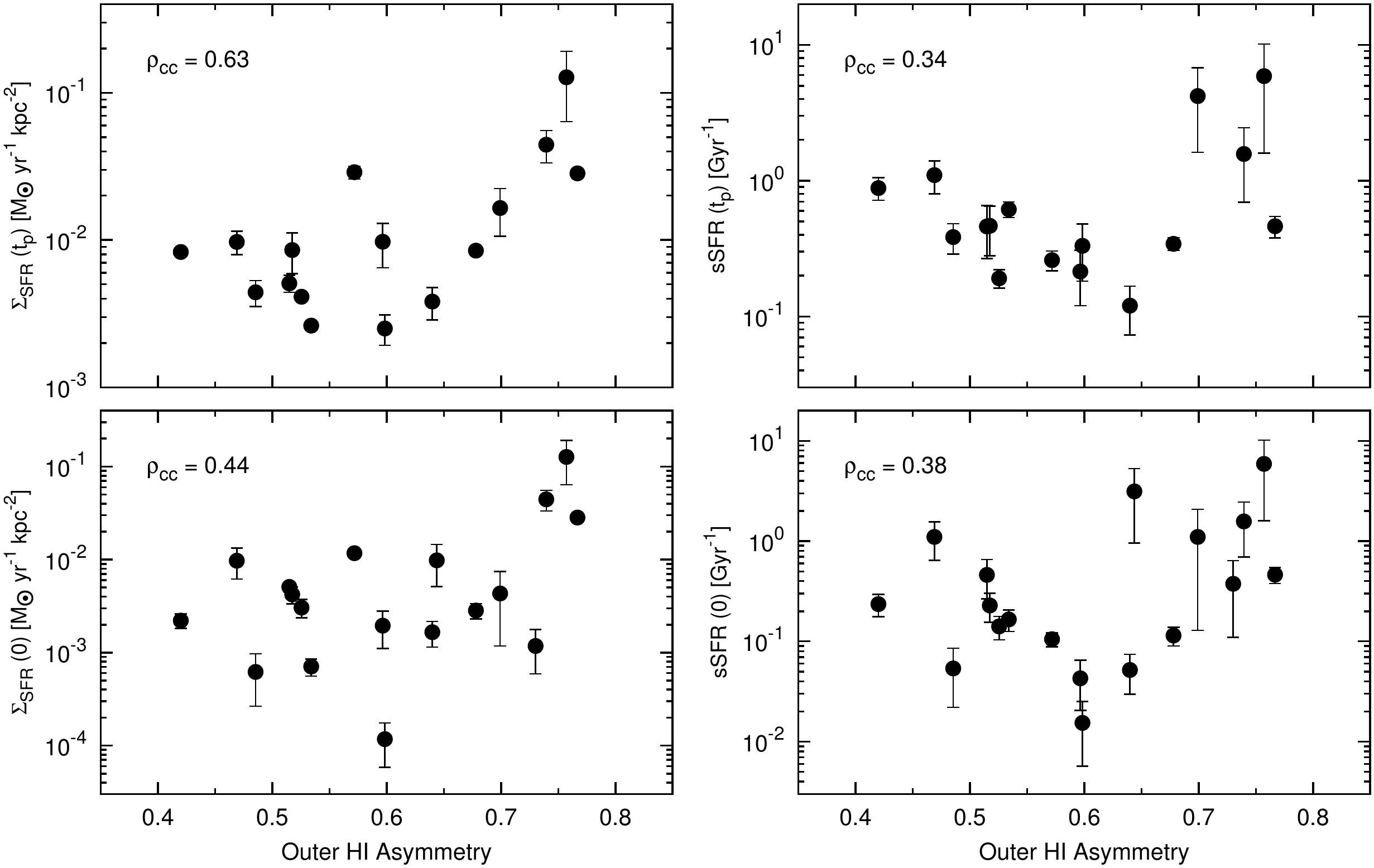}
\caption{The asymmetry parameter $A$ versus the peak SFR surface
density $\Sigma_{\rm{SFR}}(t_{\rm p})$ (\textit{top-left}), the
present-day SFR surface density $\Sigma_{\rm SFR}(0)$ (\textit{bottom-left}), 
the peak specific SFR (\textit{top-right}), and the present-day
specific SFR (\textit{bottom-right}). In each diagram, the value
of the Pearson's correlation coefficient is indicated. See
Sect.~\ref{sec:AvsSB} for details.}
\label{fig:AvsSFR}
\end{figure*}
Fig.~\ref{fig:Ahisto} shows that starburst dwarfs systematically
have \textit{higher} values of $A$ than typical Irrs. The mean and
median values of these distributions are, respectively, 0.60 and
0.60 for the sample of starburst dwarfs, and 0.47 and 0.41 for
the control sample of Irrs. Since the total \hi maps of starburst
dwarfs have larger beam-sizes than those of Irrs, the difference
between the two samples may be even larger. Given the possible
effects of beam smearing on $A$, we did not perform a statistical
analysis of the two distributions (e.g. using a Kolmogorov-Smirnoff
test). It is clear, however, that starburst dwarfs generally have
more asymmetric \hi morphologies in their outer regions than typical Irrs. 

Two galaxies from the control sample have very high values of $A$,
comparable with those of the most disturbed starburst dwarfs. These
objects are NGC~404 ($A = 0.67$) and DDO~6 ($A = 1$). NGC~404 is
at the high-mass end of the dwarf classification ($M_{\rm B} \simeq
-16.2$) and shows an unusual lenticular morphology for a dwarf
galaxy. Several authors \citep[e.g.][]{Thilker2010} argued that
NGC~404 may have experienced a merger in the last $\sim$1~Gyr
given that it has an inner, counter-rotating stellar core
\citep{Bouchard2010} and an outer, extended \hi ring \citep{DelRio2004}
hosting recent, low-level star-formation \citep{Thilker2010}.
Considering these facts, the relatively high value of $A$ is
not surprising, and demonstrates that our definition of $A$
can successfully identify past interacting/merging systems.
Regarding DDO~6, both \citet{Skillman2003} and \citet{Weisz2011}
classified this object as a ``transition'' dwarf, i.e. a low-mass
galaxy with detected \hi emission but little or no H$\alpha$
flux \citep{Mateo1998}. In DDO~6, \hi emission is detected only
on one side of the galaxy (similarly to UGC~6541 in our sample),
hence this object has an extremely high value of $A$. It would
be interesting to investigate whether this galaxy has experienced
a recent starburst. Intriguingly, the well-studied ``transition''
dwarf Antlia has been classified as a starburst by \citet{McQuinn2012a},
and has a \hi distribution similar to DDO~6 and UGC~6541 (see
\citealt{Ott2012}; Antlia is not included here because it has
$M_{\rm B} \lesssim -11$).
 
\subsection{\hi asymmetries versus starburst properties}\label{sec:AvsSB}

\begin{figure*}
\centering
\includegraphics[width=\textwidth]{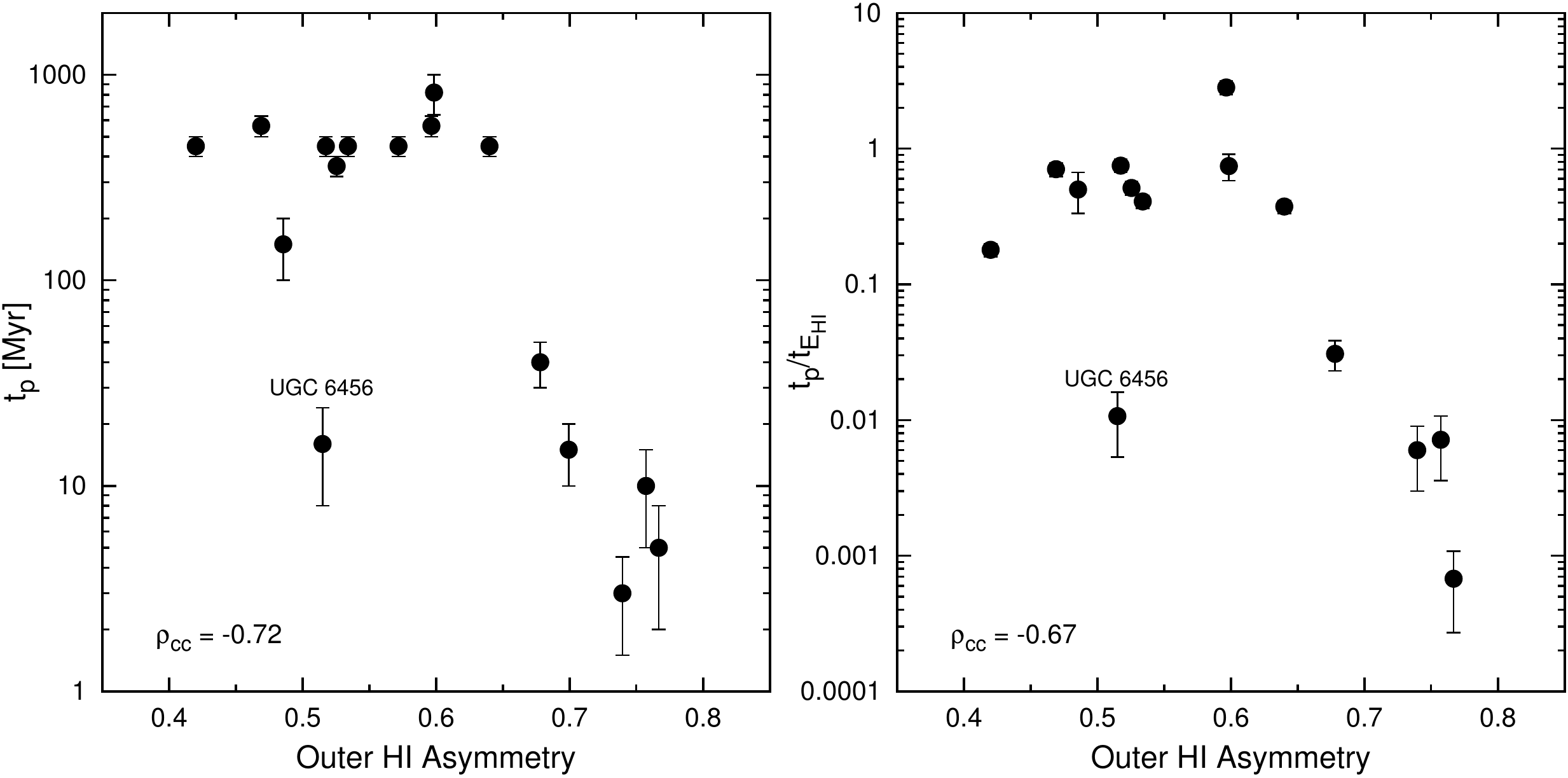}
\caption{The asymmetry parameter $A$ versus the look-back time
$t_{\rm p}$ at the peak of the SFH (\textit{left}) and the ratio
$t_{\rm p}/t_{\rm_{E_{\hiA}}}$, which roughly estimates the number
of orbits that the outermost gas has completed since the onset
of the starburst. In both panels, the value of the Pearson's
correlation coefficient is indicated. See Sect.~\ref{sec:AvsSB}
for details.}
\label{fig:AvsTp}
\end{figure*}
We now investigate the possible relations between $A$ and the
properties of the starburst as derived from the HST studies of the
resolved stellar populations. We consider the following quantities
(see Table~\ref{tab:SFRprop}):
\begin{enumerate}
\item the birthrate parameter $b=\rm{SFR_{p}}/\overline{SFR}$, where
SFR$_{\rm p}$ is the peak SFR over the past 1 Gyr and $\overline{\rm SFR}$
is the average SFR over the past 6 Gyrs \citep[see][]{McQuinn2010a};
\item the peak SFR surface density $\Sigma_{{\rm SFR}}(t_{\rm p})=
{\rm SFR}_{\rm p}/(\pi R_{\rm{opt}}^{2})$, where $R_{\rm opt}$ is
defined as 3.2 exponential scale-lengths (see LVF14);
\item the present-day SFR surface density $\Sigma_{\rm SFR}(0)={\rm
SFR_{0}}/(\pi R_{\rm opt}^{2})$, where $\rm{SFR_{0}}$ is the average
SFR over the last 10 Myr;
\item the specific SFR (sSFR) calculated both as SFR$_{0}/M_{*}$ and
as SFR$_{\rm p}/M_{*}$;
\item the look-back time $t_{\rm{p}}$ at SFR$_{\rm p}$.
\end{enumerate}
In particular, $t_{\rm p}$ can be considered as the typical ``age''
of the starburst, allowing us to distinguish between ``old'' bursts
(with $t_{\rm p}\gtrsim100$ Myr) and ``young'' bursts (with $t_{\rm p}
\lesssim100$ Myr). The SFHs of 5 galaxies (NGC~2366, NGC~4068, UGC~4483,
UGC~9128, and SBS~1415+437) show two distinct peaks with similar SFRs
(consistent within 1$\sigma$). In these cases, we consider the SFR and
the look-back time of the older peak since this is the one that formed
more stars, given that the SFR is averaged over a larger time-bin
(typically a factor of $\sim$4, see \citealt{McQuinn2010a}). For
UGC~6541 and I~Zw~36, the recent SFH is not well constrained
\citep[see][]{SchulteLadbeck2000, SchulteLadbeck2001}, thus we have
no robust estimate of $\Sigma_{{\rm SFR}} (t_{\rm p})$, sSFR$_{\rm p}$,
and $t_{\rm p}$.

In Fig.~\ref{fig:AvsSFR}, we plot $A$ versus the SFR indicators $\Sigma
_{\rm SFR}(0)$, $\Sigma_{\rm SFR}(t_{\rm p})$, sSFR(0), and sSFR($t_{\rm p}$).
To quantify possible trends in these diagrams, we calculated the Pearson's
correlation coefficient $\rho_{\rm cc}$, where $\rho_{\rm cc}=\pm1$ for an
ideal linear correlation/anticorrelation, whereas $\rho_{\rm cc}\simeq0$
if no correlation is present. We found values of $\rho_{\rm cc} \simeq 0.3$
to 0.4, except for the $A-\Sigma_{\rm SFR}(t_{\rm p})$ diagram that yields
$\rho_{\rm cc} \simeq 0.6$. We think that this weak trend is \textit{not}
significant because it is driven by 3 galaxies (NGC~4449, I~Zw~18, and NGC~1705) 
that have $t_{\rm p}\simeq10$~Myr: for these objects the values of $\Sigma
_{\rm SFR}(t_{\rm p})\simeq \Sigma_{\rm SFR}(0)$ may be systematically
enhanced with respect to galaxies with older bursts because the SFR is
averaged over a smaller time bin ($\sim$10 Myr versus $\sim$50-100 Myr),
given that the intrinsic time-resolution of the SFHs increase with
decreasing look-back time \citep[see e.g.][]{McQuinn2010a}. We also
found no clear correlation with $b$.

Fig.~\ref{fig:AvsTp} (left) shows that a clear trend is present between
$A$ and $t_{\rm p}$ ($\rho_{\rm cc}\simeq-0.7$): galaxies hosting a ``young''
burst generally have a more asymmetric \hi distribution than galaxies
hosting an ``old'' one, further suggesting a close link between the outer,
disturbed gas morphology and the central, recent starburst activity.
Galaxies with minor asymmetries ($A\lesssim 0.6$) have values of $t_{\rm p}
\simeq500$~Myr (apart from UGC~6456), which are comparable within a factor
of $\sim$2 with the orbital times $t_{\rm E_{\hi}}$ at the outermost radii
$E_{\hi}$. This is shown in Fig.~\ref{fig:AvsTp} (right), where we plot
$A$ against the ratio $t_{\rm p}/t_{\rm E_{\hi}}$. We recall that $t_{\rm
E_{\hi}}$ is an order-of-magnitude estimate, thus the ratio $t_{\rm p}
/t_{\rm E_{\hi}}$ provides only a rough measure of the number of orbits
that the outer gas may have completed since the epoch of the most intense
star-formation activity. Despite these uncertainties, Fig.~\ref{fig:AvsTp}
(right) clearly indicates that galaxies hosting an ``old'' burst may have
had enough time to complete an entire revolution around the center and,
thus, regularize their outer \hi distribution.

Finally, we discuss the ``outlier'' UGC~6456 (VII~Zw~403) indicated
in Fig.~\ref{fig:AvsTp}. This galaxy has been recently studied by
\citet{Simpson2011}, who pointed out the relatively regular \hi
morphology and the lack of a clear external trigger. The total
\hi map and velocity field of \citet{Hunter2012}, however, show
a tail/extension to the South-West. We also detected this feature
in our total \hi map, but it is below the $3\sigma$ column density
sensitivity of the observations, having $N_{\hi}\lesssim 5 \times
10^{19}$ cm$^{-2}$. Deeper \hi observations are needed to confirm
whether this \hi tail is real. The location of UGC~6456 in Figs.
\ref{fig:AvsSFR} and \ref{fig:AvsTp} would change if one adopts
column-density thresholds $\lesssim5\times 10^{19}$ cm$^{-2}$, but
unfortunately these low values cannot be consistently adopted here
due to the limited sensitivity of the \hi observations for several
galaxies in our sample.

\section{Individual galaxies and their environment}\label{sec:indi}

In the following, we discuss in detail the \hi properties of individual
galaxies in our sample and describe their nearby environment. We used
the NASA/IPAC Extragalactic Database (NED\footnote{The NASA/IPAC
Extragalactic Database (NED) is operated by the Jet Propulsion
Laboratory, California Institute of Technology, under contract with
the National Aeronautics and Space Administration.}) to search for
nearby objects with measured systemic velocities within $\pm300$~km~s$^{-1}$
with respect to the starburst dwarf. Table~\ref{tab:envir} provides
the 3 nearest galaxies to each starburst dwarf, together with their
basic properties \citep[from][]{Karachentsev2013}. We checked that
these objects are actual galaxies by visual inspection, and excluded
background/foreground galaxies when accurate, redshift-independent
distances were available. Since most of the starburst dwarfs considered
here have distances $D\lesssim 7$~Mpc, Table~\ref{tab:envir} should
be nearly complete down to dwarf galaxies with total magnitudes
$M_{\rm B} \simeq -11$ and mean surface brightnesses $\mu_{\rm B}
\simeq 25$ mag~arcsec$^{-2}$ \citep[cf. with][]{Karachentsev2004,
Karachentsev2013}. SBS~1415+437 and I~Zw~18, however, have $D\simeq14$~Mpc
and $D\simeq18$~Mpc, respectively, thus they may have faint companions
that have not been identified by optical surveys. An object with a
peculiar velocity of $\sim$200~km~s$^{-1}$ covers $\sim$200~kpc in
$\sim$1~Gyr, thus it is possible that a galaxy at a projected distance
$D_{\rm p}\lesssim200$~kpc from a starburst dwarf might have triggered
the burst by a \textit{past collision} on an hyperbolic orbit
\citep[e.g.][]{Noguchi1988}. Most of the galaxies in our sample have
such a potential perturber, except for NGC~1705, NGC~6789, and UGC~9128.
We stress, however, that the presence of such a companion does not
guarantee that the starburst has been triggered by a past collision,
given that the relative orbits of the two galaxies are unknown.

\textbf{NGC 625} has a $\sim$2 kpc \hi tail to the North-West, that
shows a coherent kinematic structure at $V_{\rm los}\simeq420$ km~s$^{-1}$.
A second tail/extension is present to the South-East, but it does not
show a clear kinematic structure. Our total \hi map and velocity field are
in close agreement with those from \citet{Cote2000} and \citet{Cannon2004}.
NGC~625 is part of the Sculptor group, but it is quite far from the
central, massive galaxy NGC~253, being at a projected distance of
$\sim$1.3 Mpc \citep{Karachentsev2005}.

\textbf{NGC 1569} has a heavily disturbed \hi distribution. A \hi cloud
with $M_{\hi}\simeq 2 \times 10^{7}$ M$_{\odot}$ lies at $V_{\rm los}
\simeq-150$ km~s$^{-1}$ to the East of the galaxy, and is connected to
the main \hi distribution by a thin bridge \citep[see also][]{Stil1998}.
The datacube is strongly affected by Galactic emission, thus the total
\hi map and the velocity field are uncertain. Our results are in close
agreement with those from \citet{Stil1998, Stil2002} and \citet{Johnson2012}.
NGC~1569 is part of the IC~432 group \citep{Grocholski2008} and has a
nearby companion (UGCA~92) at a projected distance of $\sim$70~kpc with
a similar systemic velocity (within $\sim$20 km~s$^{-1}$).

\textbf{NGC 1705} has an extended, warped \hi disk. The \hi disk shows
relatively regular morphology and kinematics, but it is strongly
offset with respect to the stellar component: the optical and kinematic
centers differ by $\sim$550 pc, while the optical and kinematic PAs differ
by $\sim$45$^{\circ}$ (see LVF14). To the North-East, there is also a
small \hi extension with peculiar kinematics, that may be associated
with the H$\alpha$ wind \citep[see][]{Meurer1998, Elson2013}. NGC~1705
appears very isolated: the two nearest objects (LSBG~F157-089 and
MRSS~157-121650) are at a projected distance of $\sim$0.5~Mpc, but may
be members of the Dorado group at $D\simeq18$~Mpc (see \citealt{Firth2006}
and \citealt{Evstigneeva2007}, respectively). Three other objects
(NGC~1533, IC~2038, and IC~2039) lie at $\sim$7$^{\circ}$ from NGC~1705,
but they seem to be background galaxies at distances of $\sim$20~Mpc
(based on the Tully-Fisher relation).

\begin{table*}
\begin{minipage}{\textwidth}
\small
\setlength{\tabcolsep}{5pt}
\caption{Environment of the starburst dwarfs in our sample. We used information
from NED to calculate the projected distances $D_{\rm p}$ from the nearest galaxies
and the difference between their respective systemic velocities $\Delta V_{\rm sys}$.
The properties of the nearest galaxies are taken from \citet{Karachentsev2013},
and given only when their distances are estimated from the TRGB (the distance
of NGC~2403 is estimated from Cepheids). Objects without an accurate distance
estimate may be background/foreground galaxies. The properties of I~Zw~18~C are
taken from \citet{Lelli2012a}. The morphological types are taken from NED and/or
\citet{Karachentsev2013}.}
\begin{center}
\begin{tabular}{l l l c c c c c c c c}
\hline
Galaxy   & Membership     & Nearest galaxies & $D_{\rm p}$ & $\Delta V_{\rm sys}$ & Type & Dist & $M_{\rm B}$ & $\log(M_{\hi})$ & $W_{50,\, \hiA}$   \\
         &                &                  & (kpc)       & (km~s$^{-1}$)        &      & (Mpc)& (mag)       & ($M_{\odot}$)   & (km~s$^{-1}$) \\
\hline
NGC 625  & Sculptor group & ESO245-005       & 203         & $-7$                 & Im   & 4.4   & -15.6       & 8.58            & 60  \\
         & (periphery)    & CFC97 Sc 24$^\alpha$ & 402  & -236       & ...  & ...   & ...         & ...             & 92  \\
         &                & GSD 106          & 619         & 132                  & ...  & ...   & ...         & ...             & ... \\
\rule{0pt}{3.3ex}
NGC 1569 & IC~432 group   & UGCA 92          & 74          & 19                   & Irr  & 3.0   & -15.6       & 8.17            & 56  \\     
         &                & Cam B            & 190         & 157                  & Irr  & 3.3   & -11.9       & 7.08            & 21  \\
         &                & UGCA 86          & 231         & 147                  & Im?  & 3.0   & -17.9       & ...             & ... \\
\rule{0pt}{3.3ex}
NGC 1705 & Field          & LSBG F157-089$^\beta$      & 518  & 175  & ...  & ...   & ...         & ...             & ... \\
         &                & MRSS 157-121650$^\gamma$ & 562  & -50     & ...  & ...   & ...         & ...             & ... \\
         &                & SGC 0409.0-5638  & 631         & 242                  & Irr  & ...   & ...         & ...             & ... \\
\rule{0pt}{3.3ex}
NGC 2366 & M81 group      & NGC 2363$^\delta$  & 2          & -33                  & Irr  & ...   & ...         & ...             & ... \\
         & (periphery)    & UGCA~133$^\epsilon$& 133         & 110                  & Sph  & 3.2   & -12.1       & $<6.0$          & ... \\
         &                & NGC 2403           & 206         & 30                   & Scd  & 3.2   & -19.2       & 9.48            & 240 \\
\rule{0pt}{3.3ex}
NGC 4068 & CVn~I cloud    & MCG +09-20-131   & 135         & -47                  & Irr  & 4.6   & -13.1       & 7.37            & 27  \\
         &                & ASK 185765.0$^\varepsilon$  & 143 & 291     & ...  & ...   & ...         & ...             & ... \\
         &                & UGC 7298         & 145         & -33                  & Irr  & 4.2   & -12.3       & 7.28            & 21 \\
\rule{0pt}{3.3ex}
NGC 4163 & CVn~I cloud    & MCG +06-27-017$^\zeta$ & 27 & 181  & Im   & 4.8   & -13.0       & ...             & ... \\
         &                & NGC 4190         & 29          & 70                   & Im/BCD & 2.8 & -13.9       & 7.46            & 49  \\
         &                & DDO 113          & 30          & 126                  & Sph? & 2.9   & -11.5       & $<5.56$         & 32 \\
\rule{0pt}{3.3ex}
NGC 4214 & CVn~I cloud    & DDO 113          & 8           & -7                   & Sph? & 2.9   & -11.5       & $<5.56$         & ... \\
         &                & NGC 4190         & 23          & -63                  & Im/BCD & 2.8 & -13.9       & 7.46            & 49 \\
         &                & NGC 4163         & 34          & -133                 & BCD  & 3.0   & -13.8       & 7.16            & 32 \\
\rule{0pt}{3.3ex}
NGC 4449 & CVn~I cloud    & DDO 125          & 44          & -15                  & Im   & 2.7   & -14.3       & 7.48            & 27 \\
         &                & MCG +07-26-012   & 77          & 226                  & Im   & ...   & ...         & ...             & ...\\
         &                & DDO 120          & 155         & 252                  & Im   & ...   & ...         & ...             & ...\\
\rule{0pt}{3.3ex}
NGC 5253 & M83 group      & ESO444-084$^\eta$       & 54 & 116      & Sc   & ...   & ...         & ...             & ...\\
         &                & NGC 5264         & 108         & 68                   & Im   & 4.5   & -15.9       & 7.65            & 35 \\
         &                & HIDEEP J1337-33  & 111         & 181                  & Irr  & 4.4   & -11.1       & 6.67            & 20 \\
\rule{0pt}{3.3ex}
NGC 6789 & Local Void     & ABELL 2312:[MPC97]~04$^\theta$ & 297 & 59 & ...  & ...    & ...         & ...             & ...\\
         &                & UGC 11411        & 400         & 220                  & BCD  & ...   & ...         & ...             & ...\\
         &                & LEDA 166193      & 578         & 290                  & Irr  & ...   & ...         & ...             & 28 \\
\rule{0pt}{3.3ex}
UGC 4483 & M81 group      & M81 Dwarf A      & 92          & -45                  & Irr  & 3.5   & -11.5       & 7.06            & 21 \\
         &                & Holmberg II      & 100         & -16                  & Im   & 3.4   & -16.7       & 8.61            & 64 \\
         &                & DDO 53           & 201         & -138                 & Irr  & 3.6   & -13.4       & 7.62            & 30 \\
\rule{0pt}{3.3ex}
UGC 6456 & M81 group      &CGCG 351-049$^\iota$ & 151 & 8            & ...  & ... & ...         & ...             & ...\\
         & (periphery)    & UGC 8245$^\iota$    & 358 & 172          & Im   & ... & ...         & ...             & ...\\
         &                & DDO 82           & 683         & 158                  & Im   & 4.0   & -14.7       & ...             & ...\\
\rule{0pt}{3.3ex}
UGC 6541 & CVn~I cloud    & ASK 184683.0     & 213         & 208                  & ...  & ...   & ...         & ...             & ... \\
         & (periphery)    & ASK 185765.0$^\varepsilon$ & 289 & 247      & ...  & ...   & ...         & ...             & ...\\
         &                & NGC 3741         & 292         & -21                  & Im/BCD & 3.0 & -13.1       & 7.88            & 83 \\
\rule{0pt}{3.3ex}
UGC 9128 & Field          & LSBG F650-01$^\kappa$ & 353 & -167          & ...  & ...   & ...       & ...             & ...\\
         &                & MAPS O-383-0548118$^\lambda$ & 359 & -93    & ...  & ...   & ...       & ...             & ...\\
         &                & SDSS J145657.7+221315$^\lambda$ & 365 & -102 &...  & ...   & ...       & ...             & ...\\
\rule{0pt}{3.3ex}
UGCA 290 & NGC~4631 group?& UGC 7719         & 83          & 210                  & Sdm  & ...   & ...         & ...             & 57 \\
         &                & IC 3687          & 115         & 114                  & Im   & 4.6   & -14.6       & 7.90            & 36 \\
         &                & BTS 142          & 122         & 251                  & Irr  & ...   & ...         & ...             & 23 \\
\rule{0pt}{3.3ex}
I Zw 18  & Field          & I Zw 18 C        & 2           & -16                  & Irr  & 18.2  & -12.1       & $\lesssim8.08$  & $\sim$45\\
         &                & ASK 153750.0     & 639         & 243                  & ...  & ...   & ...         & ...             & ...\\
         &                & MGC +09-16-029   & 1052        & -159                 & ...  & ...   & ...         & ...             & ...\\
\hline
\end{tabular}
\end{center}
\label{tab:envir}
\end{minipage}
\end{table*}

\begin{table*}
\begin{minipage}{\textwidth}
\small
\setlength{\tabcolsep}{5pt}
\begin{center}
\contcaption{}
\begin{tabular}{l l l c c c c c c c c}
\hline
Galaxy   & Membership     & Nearest galaxies & $D_{\rm p}$ & $\Delta V_{\rm sys}$ & Type & Dist & $M_{\rm B}$ & $\log(M_{\hi})$ & $W_{50,\, \hiA}$   \\
         &                &                  & (kpc)       & (km~s$^{-1}$)        &      & (Mpc)& (mag)       & (10$^{7}$)      & (km~s$^{-1}$) \\
\hline
I Zw 36  & CVn~I cloud    & UGC 7639$^\mu$ & 117         & 105       & Im   & ...   & ...         & ...             & ...\\
         &                & NGC 4248$^\nu$ & 183         & 207       & ...  & ...   & ...         & ...             & ...\\
         &                & MAPS O-171-0165792 & 214       & 195                  & ...  & ...   & ...         & ...             & ...\\
\rule{0pt}{3.3ex}
SBS 1415+437 & Field      & MAPS O-221-0093662 & 179       & 105                  & ...  & ...   & ...         & ...             & ... \\
             &            & ASK 310753.0     & 492         & 18                   & ...  & ...   & ...         & ...             & ... \\
             &            & NGC 5608         & 493         & 47                   & Im   & ...   & ...         & ...             & ... \\
\hline
\end{tabular}
\end{center}

\medskip
$^\alpha$ According to \citet{Karachentsev2004}, this galaxy is
not in the Sculptor group but lie outside the Local Volume. The
value of $W_{50, \, \hi}$ is taken from \citet{Cote1997}.\\
$\beta$ This galaxy may be in the Dorado group ($D \simeq 17$~Mpc,
\citealt{Firth2006}).\\
$^\gamma$ This object may be an ultra-compact dwarf in the Dorado
group ($D \simeq 17$~Mpc, \citealt{Evstigneeva2007}).\\
$^\delta$ It is unclear whether this object is part of NGC~2366
or a separate galaxy (see Fig.~\ref{fig:n2366}).\\
$^\epsilon$ The systemic velocity of this galaxy is not reported
by NED. We used the value given by \citet{Karachentsev2013}.\\
$^\varepsilon$ \citet{Trentham2001} included this galaxy in their study
of the Ursa Major cluster. However, ASK~185765.0 is probably not a
cluster member, given that its systemic velocity is $\sim$497~km~s$^{-1}$
(from the Sloan Digital Sky Survey).\\
$^\zeta$ This galaxy is close to NGC~4163 on the sky, but inhabits
a further region of the CVn~I cloud having $D \simeq 4.8$ Mpc
(from the TRGB).\\
$^\eta$ This edge-on spiral is not a member of the M83 group and
probably is a background galaxy.\\
$^\theta$ NED classifies this object as a galaxy. It is projected on
the sky near the galaxy cluster ABELL~2312 \citep{Maurogordato1997},
but its systemic velocity indicates that it is a nearby object. In our
opinion, it is unclear whether this is a galaxy or a Galactic object.\\
$^\iota$ This object is not associated to the M81~group and may be
a background/foreground galaxy.\\
$^\kappa$ According to NED, this object may be a planetary nebula.\\
$^\lambda$ This object has also been classified as a X-ray source and a star.\\
$^\mu$ This object may have $D \simeq 7.1$~Mpc (from surface brightness
fluctuations, \citealt{Karachentsev2013}) and be outside the CVn~I cloud.\\
$^\nu$ This object may have $D \simeq 7.4$~Mpc (from the Tully-Fisher
relation, \citealt{Karachentsev2013}) and be outside the CVn~I cloud.
\end{minipage}
\end{table*}
\begin{figure*}
\centering
\includegraphics[width=\textwidth]{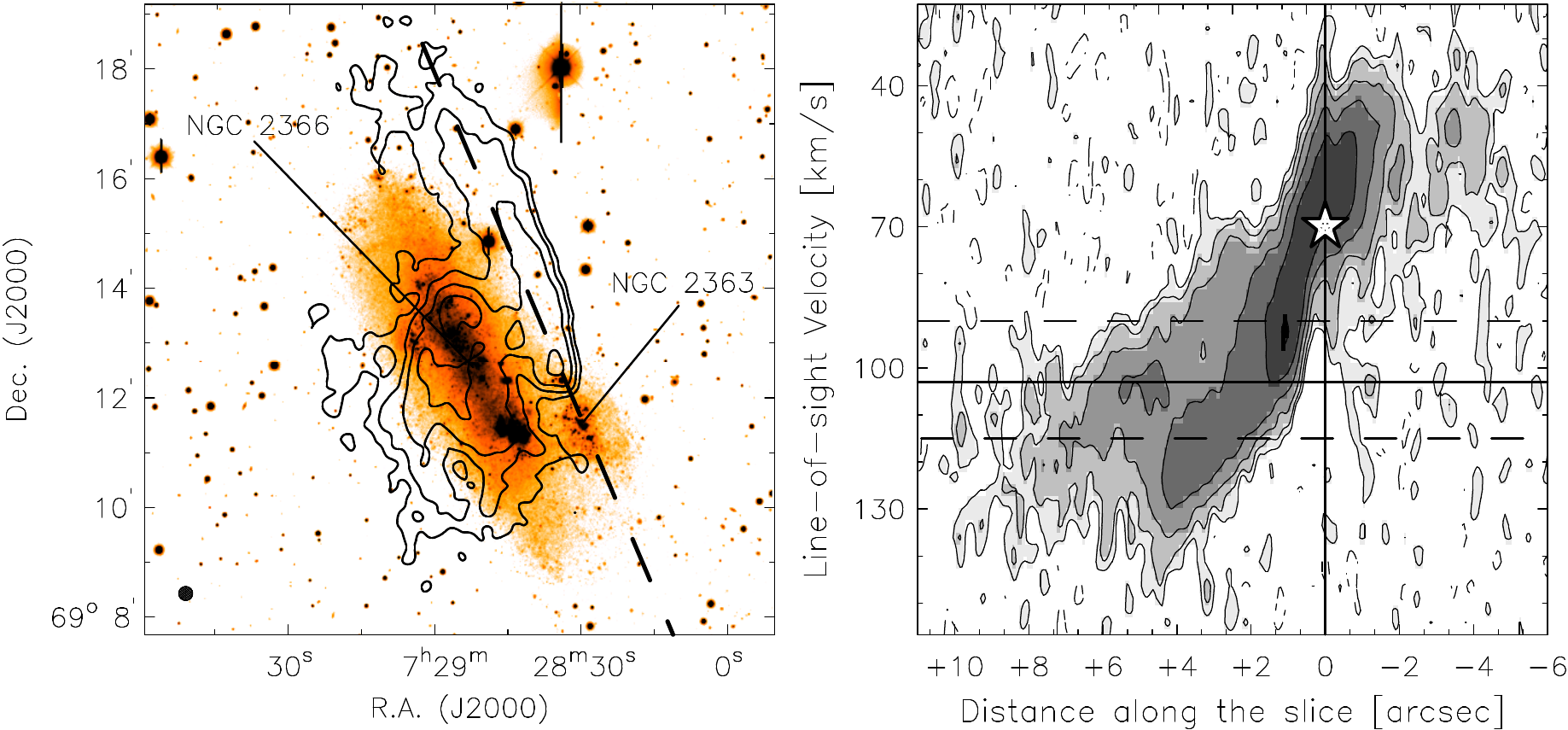}
\caption{Kinematically-anomalous gas in NGC~2363/NGC~2366. \textit{Left}:
$V$-band image overlaid with the \hi emission at 15$''$ resolution,
integrated over a narrow velocity range near the systemic velocity of
NGC~2366 (from $\sim$90 to $\sim$115 km~s$^{-1}$). The \hi column
densities are 2, 4, 8, 16 $M_{\odot}$~pc$^{-2}$. The arrows indicate
the centers of NGC~2363 and NGC~2366. The circle to the bottom-left
shows the \hi beam. \textit{Right}: Position-Velocity diagram taken
along the dashed line shown in the left-panel. The solid, horizontal
line indicates the systemic velocity of NGC~2366 (103 km~s$^{-1}$),
while the dashed, horizontal lines show the velocity range that
has been used to create the \hi map in the left-panel. The vertical
line corresponds to the spatial position of NGC~2363, while the
star shows its optical systemic velocity (70 km~s$^{-1}$), as
given by NED. Contours are at -1.5 (dashed), 1.5, 3, 6, and 12
$\sigma$, where $\sigma = 0.66$ mJy/beam.}
\label{fig:n2366}
\end{figure*}

\textbf{NGC 2366} has a \hi disk with a broad extension to the South-East
and a strong kinematic distortion to the North-West. Fig.~\ref{fig:n2366}
(left) shows an optical image overlaid with the \hi emission at 15$''$
resolution, integrated over a narrow velocity range near the systemic
velocity (between $\sim$90 and $\sim$115 km~s$^{-1}$). The gas to the
North-West does not follow the rotation of the \hi disk \citep[see also
Fig.~3 of][]{Oh2008} and may be associated with the secondary star-forming
body to the South-West (NGC~2363). Fig~\ref{fig:n2366} (right) shows a
Position-Velocity diagram taken along the dashed-line in Fig.~\ref{fig:n2366}
(left). Intriguingly, the PV-diagram displays a steep velocity gradient
coinciding with the spatial position and optical systemic velocity of
NGC~2363 (indicated by the star). This may indicate that this velocity
gradient is due to rotation in a local potential well. However, given
the overall rotation of the \hi disk of NGC~2366, this conclusion is
uncertain. The NGC~2363/NGC~2366 system probably is an on-going minor
merger.

\textbf{NGC 4068} shows a broad \hi extension to the South-East. This
object is in the Canes~Venatici~I (CVn~I) cloud, which is an extended,
loose group inhabited by low-mass galaxies \citep{Karachentsev2003}.

\textbf{NGC 4163} shows a \hi tail to the West and, possibly, a second
tail to the South \citep[see also][]{Hunter2012}. NGC~4163 is in the
CVn~I cloud and lie close to several other Irrs (at $D_{\rm p}\simeq$30~kpc),
including the starburst dwarf NGC~4214 and the ``compact'' irregular
NGC~4190 (UGC~7232). Intriguingly, NGC~4190 has been classified as a
BCD by \citet{Karachentsev2013} and shows a disturbed \hi morphology
\citep[see][]{Swaters2002a}.

\textbf{NGC 4214} has a \hi disk with a well-defined spiral pattern.
The \hi disk is strongly warped (see LVF14) and slightly more extended
to the North-West. NGC~4214 is in the CVn~I cloud and has a small
companion galaxy (DDO~113) at a projected distance of $\sim$8~kpc.
DDO~113 likely is a gas-poor spheroidal \citep{Kaisin2008}. This object,
indeed, is within the field-of-view of the VLA but no \hi emission is
detected within the covered velocity range.

\textbf{NGC 4449} has an extremely extended \hi distribution characterized
by long filaments with column densities of $\sim$1~M$_{\odot}$~pc$^{-2}$
\citep[][]{Hunter1998}. A tidally-disturbed companion galaxy is present
towards the South-East, but it does not spatially coincide with any
gaseous feature \citep{Delgado2012}, thus its relation with the
outer \hi distribution is unclear. NGC~4449 is one of the most
massive galaxies in the CVn~I cloud \citep{Karachentsev2005}.

\textbf{NGC 5253} has a $\sim$4~kpc \hi tail to the North at $V_{\rm los}
\simeq400$ km~s$^{-1}$. Our total \hi map at $40'' \times 40''$ resolution
is slightly different from that of \citet{LopezSanchez2012} at $57.8 \times
37.5''$ resolution because we used a Gaussian-smoothed, robust-weighted
datacube instead of a natural-weighted datacube. The former cube has a
much more regular noise structure than the latter one, providing a better
estimate of the 3$\sigma$ column density sensitivity. NGC~5253 is in the
CenA/M83 group; its projected distance from the spiral galaxy M83 is
$\sim$150~kpc \citep{Karachentsev2005}.

\textbf{NGC 6789} has a regularly-rotating \hi disk with several
asymmetric features in the outer parts. This galaxy is in the 
Local Void and its nearest massive companion (NGC~6946) is at a
projected distance of $\sim$2.5~Mpc \citep{Drozdovsky2001}.

\textbf{UGC 4483} has a regularly-rotating \hi disk with a small
extension to the North-West. This galaxy is in the M81 group and
lies between the group center and the NGC~2403 sub-group
\citep{Karachentsev2002}.

\textbf{UGC 6456} has a \hi disk that is slightly more extended to the
South. The data are affected by Galactic emission, making the total
\hi map uncertain. Our results are in agreement with those of \citet[]
[]{Simpson2011}. UGC~6456 lies in the periphery of the M81 group
\citep{Karachentsev2005}.

\textbf{UGC 6541} has a strongly asymmetric \hi distribution. Gas emission
is detected only in the Northern half of the galaxy. UGC~6541 is located
to the North-Western edge of the CVn~I cloud \citep{Karachentsev2003}.
Another BCD (NGC~3741, \citealt{Karachentsev2013}) lies at a projected
distance of $\sim$300~kpc.

\textbf{UGC 9128} has a relatively regular \hi distribution, but the
optical and kinematic position angles differ by $\sim$30$^{\circ}$.
This galaxy appears very isolated; the closest massive galaxy is the
Milky Way at $D\simeq2.2$~Mpc \citep{Karachentsev2013}.

\textbf{UGCA~290} has a peculiar \hi distribution that is off-set with
respect to the stellar component. Our total \hi map at $20'' \times 20''$
resolution is less extended than the one obtained by \citet{Kovac2009}
using WSRT data at $52.2''\times30.9''$ resolution, but the \hi fluxes
are consistent within the uncertainties, indicating that our total \hi
map is not missing diffuse \hi emission. UGCA~290 may be part of the
NGC~4631 group; its projected distance from NGC~4631 is $\sim$700~kpc.

\begin{figure}
\centering
\includegraphics[width=0.5\textwidth]{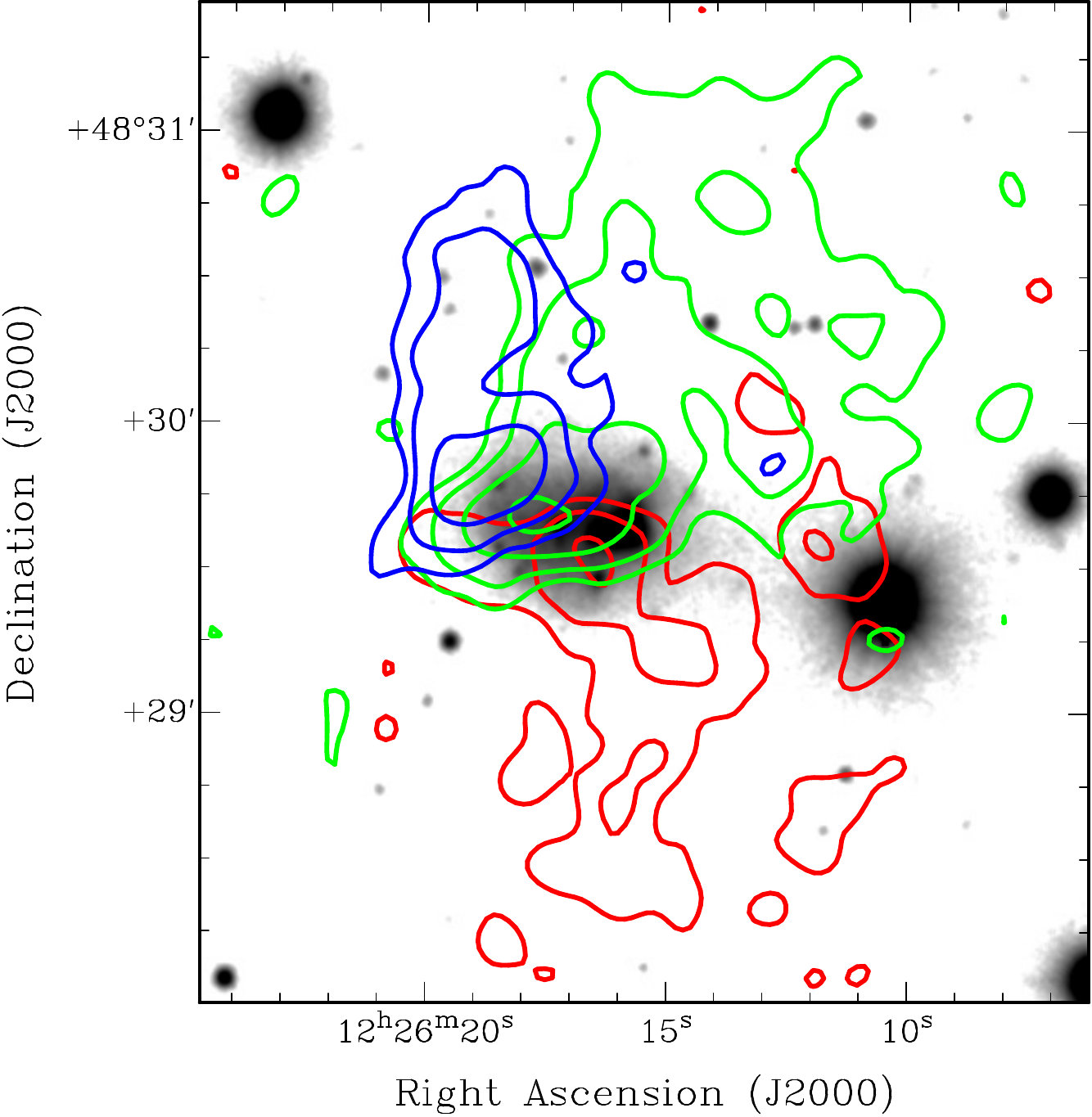}
\caption{\hi emission in I~Zw~36 at 10$''$ resolution, integrated
across 3 velocity ranges: 250 to 260 km~s$^{-1}$ (blue),  275 to
285 km~s$^{-1}$ (green), and 300 to 310 km~s$^{-1}$ (red).
Contours are at 1.2, 2.4, 4.8, and 9.6 M$_{\odot}$~pc$^{-2}$.}
\label{fig:Izw36}
\end{figure}
\textbf{I~Zw~18} has been studied in detail by \citet{Lelli2012a}.
The total \hi map presented here is slightly different from that in
\citet{Lelli2012a} because it was constructed using a mask at 60$''$
resolution (instead of 45$''$ resolution) for consistency with the
other galaxies. The most likely interpretation of this system is an
interaction/merger between two (or more) gas-rich dwarfs.

\textbf{I~Zw~36} has an extended, asymmetric \hi distribution,
that is kinematically-connected to a central rotating disk (see
\citealt{Ashley2013} and LVF14). Data at 10$''$ resolution (see
Fig.~\ref{fig:Izw36}) reveal that the \hi emission forms a
tail-like structure to the South at receding velocities
($V_{\rm los}\simeq300$ to 310 km~s$^{-1}$; $V_{\rm sys} = 277$
km~s$^{-1}$) and a broad extension to the North near the systemic
velocity ($V_{\rm los}\simeq270$ to 290 km~s$^{-1}$), possibly
connected to the approaching side of the disk ($V_{\rm los}\simeq250$
to 260 km~s$^{-1}$). There are no optical features associated with
the extended gas down to $\mu_{\rm R} \simeq 26$ mag~arcsec$^{-2}$.
I~Zw~36 is in the CVn~I cloud.

\textbf{SBS~1415+437} has an extended, lopsided \hi disk. The
galaxy is at a relatively large distance ($\sim$13.6 Mpc),
thus it is possible that faint, nearby companions have not
been identified by optical surveys.

\section{Discussion}

In Sect.~\ref{sec:morpho} we found that starburst dwarfs show
a large variety of \hi morphologies. Several of them have
heavily disturbed \hi morphologies, characterized by strong
asymmetries, long filaments, and/or large offsets between the
stellar and \hi distributions. Other starburst dwarfs, instead,
show minor asymmetries, characterized by \hi extensions and/or
small tails in the outer parts. In Sect.~\ref{sec:asym} we
introduced the parameter $A$, quantifying the outer \hi asymmetry,
and measured it for both our sample of starburst dwarfs and
a control sample of typical Irrs. We found that starburst
dwarfs systematically have more asymmetric \hi morphologies
that typical Irrs, although there is a ``grey area'' for
$0.4 \lesssim A \lesssim 0.5$ where one can find both starburst
and non-starburst dwarfs with lopsided \hi morphologies.
Lopsidedness is a common phenomenon among spirals and irregular
galaxies \citep[e.g.][]{Baldwin1980, Verheijen2001, Swaters2002a},
and it has been suggested that it may be due to past
interactions and/or accretion events \citep[e.g.][]{Sancisi2008}.
Alternatively, one may conceive that the outer \hi asymmetries
are the result of gaseous outflows due to stellar feedback.
This latter hypothesis seems unlikely, as we now discuss.

Hydrodynamical simulations show that gaseous outflows
generally follow the path of least resistance from the ISM
\citep{MacLow1999, Dercole1999, Cooper2008} and, thus, develop
perpendicularly to the galaxy major axis. Several starburst
dwarfs do show diffuse H$\alpha$ emission roughly perpendicular
to the galaxy major axis \citep[see e.g. Fig.~7 of][]{Lee2009},
which likely traces an outflow (although the H$\alpha$ gas 
generally does \textit{not} escape from the galaxy potential
well; e.g. \citealt{Martin1996, Martin1998, vanEymeren2009b,
vanEymeren2009a, vanEymeren2010}). The \hi gas in the outer
regions, instead, often has a tail-like morphology and does not
show any preferential direction with respect to the galaxy major
axis (see Fig.~\ref{fig:mosaic1}). Moreover, in general
there is little (if any) correlation between the \hi and
H$\alpha$ emission in the outer galaxy regions; see e.g.
I~Zw~18 \citep[Fig.~9 of][]{Lelli2012a} and NGC~1705
\citep[Fig.~8 of][]{Elson2013}. We also note that, if the
outer \hi asymmetries were due to stellar feedback, one
may expect a correlation between the asymmetry parameter
$A$ and some SFR indicators, given that a higher star-formation
activity would produce stronger outflows and, thus, more
asymmetric \hi distributions. We found no convincing correlation
between $A$ and either the SFR surface density or the specific
SFR (see Sect.~\ref{sec:AvsSB}). For all these reasons, we think
that the \hi emission in the outer regions is not due to gas
outflows, but it indicates that the starburst is triggered by
\textit{external mechanisms}, such as interactions/mergers
between gas-rich dwarfs or cold gas accretion from the IGM.
\begin{table*}
\centering
\caption{Other starburst dwarfs with high-quality \hi observations.}
\small
\begin{tabular}{l l l}
\hline
Name(s)            & Properties                                         & References \\
\hline
II~Zw~33 (Mrk 1039)& LSB companion II~Zw~33~B                           & \citet{Walter1997}\\
HS~0822+3542       & LSB companion SAO 0822+3542                        & \citet{Chengalur2006}\\
Mrk~324 (UGCA 439) & LSB companion EXG~0123-0040; \hi tail?            & \citet{vanZee2001}\\
NGC~2537 (Mrk 86)  & Companion IC~2233 (Sd); outer \hi arm             & \citet{Matthews2008}\\
UM~461/UM~462      & BCD pair with lopsided \hi disks                  & \citet{vanZee1998b}\\
Haro 4 (Mrk 26)    & Possible interaction with NGC~3510 (Sm)            & \citet{BravoAlfaro2004}\\
Mrk 108 (IC 2450)  & Interacting with NGC~2814 (Sb) $\&$ NGC~2820 (Sc)  & \citet{Kantharia2005}\\
II Zw 70/ II Zw 71 & Interacting pair                                   & \citet{Cox2001}\\
SBS 0335-052       & Interacting pair                                   & \citet{Pustilnik2001a, Ekta2009}\\
SBS 1129+576/577   & Interacting pair                                   & \citet{Ekta2006}\\
II~Zw~40 (UGC 116) & Advanced merger                                    & \citet{vanZee1998b}\\
IC~10 (UGC 192)    & Long \hi filaments and plumes                     & \citet{Manthey2008}\\
Haro~36 (UGC 7950) & \hi filament                                      & \citet{Ashley2013}\\
Mrk~1418 (UGC 5151)& \hi plumes and clouds                             & \citet{vanZee2001}\\
FCC~35             & High-velocity \hi complex                         & \citet{Putman1998}\\
Haro 2 (Mrk 33)    & \hi extension and small tail/cloud                & \citet{Thuan2004}\\
NGC~4861 (UGC 8098)& \hi disk extended towards a \hi cloud            & \citet{Thuan2004}\\
Mrk~900 (NGC 7077) & Lopsided \hi disk                                 & \citet{vanZee2001}\\
Mrk~750            & Lopsided \hi disk                                 & \citet{vanZee2001}\\
UM~439 (UGC 6578)  & Lopsided \hi disk                                 & \citet{vanZee1998b}\\
UM~323             & Lopsided \hi disk, possibly warped                & \citet{vanZee2001}\\
UM~38              & Relatively regular \hi disk                       & \citet{vanZee2001}\\
NGC~2915           & Extended, warped \hi disk                         & \citet{Meurer1996, Elson2010}\\
\hline
\end{tabular}
\label{tab:lit}
\end{table*}

In Sect.~\ref{sec:AvsSB} we found that there is a significant
correlation between $A$ and the look-back time at the peak of
the star-formation activity $t_{\rm p}$ (see Fig.~\ref{fig:AvsTp},
left). Galaxies hosting an ``old'' burst ($\gtrsim$100 Myr)
have low values of $A$, while galaxies hosting a ``young''
burst ($\lesssim$100 Myr) have a progressively more asymmetric
\hi distribution. In particular, galaxies with lopsided \hi
morphologies ($A\lesssim 0.6$) have values of $t_{\rm p}\simeq
500$~Myr that are comparable with the orbital times in the
outer regions (see Fig.~\ref{fig:AvsTp}, right). This
suggests that the \textit{differential} rotation in the outer
galaxy regions could have had enough time to partially
regularize the \hi distribution since the epoch of the
interaction/accretion event that possibly triggered the
starburst. In particular, galaxies with extended, strongly-warped,
and regularly-rotating \hi disks, such as NGC~4214 (LVF14)
and NGC~2915 \citep{Elson2010}, may represent an advanced
stage of the interaction/accretion phenomenon, as it has
already been suggested by \citet{Sancisi2008}. On the other
hand, a galaxy like NGC~1705, which has a warped \hi disk
that is strongly off-set with respect to the stellar component
(see LVF14), may be in an earlier stage where the outer \hi
gas is still in the process of settling down. This is in
agreement with the very recent starburst activity ($t_{\rm p}
\simeq3$~Myr) observed in this galaxy \citep{Annibali2003}.

Recent \hi studies by \citet{Ekta2008}, \citet{Ekta2010a}, and
\citet{LopezSanchez2010b} have also highlighted the importance
of interaction/accretion events in triggering the starburst in 
low-mass galaxies. In Table~\ref{tab:lit}, we list further examples
of starburst dwarfs with high-quality \hi observations. This list
is by no means complete. We have, however, carefully inspected
the published total \hi maps and velocity fields of these galaxies,
and report their main properties in Table~\ref{tab:lit}. These
galaxies do \textit{not} have accurate SFHs from HST observations,
but are thought to be experiencing a starburst based on their blue
colors, high surface brightnesses, and/or strong emission lines.
We also have no direct information on the ``age'' of the starburst.
However, considering the observed trend between \hi asymmetry and
$t_{\rm p}$, we are probably observing starburst dwarfs at different
stages of the interaction/accretion process. In particular, we
distinguish between four main ``classes'' or ``evolutionary stages'':
\begin{enumerate}
 \item Starburst dwarfs that have a nearby companion ($\lesssim200$~kpc)
but show \textit{no} sign of strong interactions, such as \hi bridges
or tails (e.g. II~Zw~33, \citealt{Walter1997}). These systems may either
be experiencing a mild tidal interaction or represent a late stage after
a fly-by.
 \item Starburst dwarfs that are clearly interacting with a companion 
(e.g. II~Zw~70/II~Zw~71, \citealt{Cox2001}) or are in an advanced stage
of merging (e.g. II~Zw~40, \citealt{vanZee1998b}).
 \item Starburst dwarfs that are relatively isolated and show a
 heavily-disturbed \hi morphology (e.g. IC~10, \citealt{Manthey2008}),
 which may be due to a recent interaction/merger or cold gas accretion
 from the environment.
 \item Starburst dwarfs that are relatively isolated and have an
 extended, lopsided \hi disk (e.g. UM~439, \citealt{vanZee1998b})
 or a pronounced warp (e.g. NGC~2915, \citealt{Elson2010}).
\end{enumerate}
Our galaxy sample includes starburst dwarfs from all these 4 classes.
As we described in Sect.~\ref{sec:indi}, NGC~4214 and NGC~4163 have
several nearby companions belonging to the CVn~I cloud and, thus, fit
into class (i). \citet{Grocholski2008} argued that NGC~1569 and UGCA~290
form a pair of galaxies in the IC~432 group similar to the LMC and
the SMC in the Local Group; in this case, NGC~1569 would also belong
to class (i). There may be more starburst dwarf in this class, having
galaxies at projected distances $D_{\rm p}\lesssim 200$~kpc and
differences in their systemic velocities $\lesssim 300$~km~s$^{-1}$,
but the lack of accurate distance estimates for their potential
companions prevents us from unambiguously classifying them (see
Sect.~\ref{sec:indi}). I~Zw~18, NGC~4449, and NGC~2366 are probably
undergoing a minor merger (see \citealt{Lelli2012a},
\citealt{Delgado2012}, and Sect.~\ref{sec:indi}, respectively)
and, thus, belong to class~(ii). I~Zw~36, UGC~6431, UGCA~290, and
NGC~625 can be included in class~(iii), whereas UGC~4483, UGC~6456,
UGC~9128, and SBS~1415+437 belong to class~(iv). NGC~6789 and
NGC~5253 are somewhat intermediate between class (iii) and (iv),
having $A\simeq 0.6$.

The observational evidence presented so far indicates that past
and on-going interaction/accretion events play an important
role in triggering the starburst in low-mass galaxies. Moreover,
interaction/mergers between gas-rich dwarfs may provide the
mechanism that forms the central concentration of mass observed
in starburst dwarfs \citep{Lelli2012a, Lelli2012b, Lelli2014a}.
Numerical simulations, indeed, indicate that interactions/mergers
between gas-rich dwarfs can lead to an overall contraction of
the disk and form a central mass concentration \citep[e.g.]
[]{Bekki2008}. We stress, however, that several galaxies in
our sample show remarkably symmetric optical morphologies (down
to $\sim$26 mag arcsec$^{-2}$), whereas the \hi distribution is
heavily perturbed (see e.g. I~Zw~36 in Fig.~\ref{fig:Izw36}).
To unambiguously identify galaxy interaction as the main
triggering mechanism, one would need deep optical observations
(down to $\sim29-30$ mag~arcsec$^{-2}$) to search for stellar
tidal features associated with the \hi features. In the case
that stellar tidal features would still remain undetected,
the remaining possibility is that starburst dwarfs are directly
accreting gas from the IGM. Cold flows of gas are predicted by
$\Lambda$CDM models of galaxy formation \citep{Keres2005,
Dekel2006}. In particular, \citet{Keres2005} argued that these
cold flows might still take place at $z\simeq0$ in low-mass
galaxies residing in low-density environments. As we discussed
in Sect.~\ref{sec:indi}, most starburst dwarfs in our sample
inhabit similar environments as typical Irrs, such as galaxy
groups and small associations. Thus, it is unclear why cosmological
cold flows would be visible only in starburst dwarfs and not
in typical Irrs, unless they are highly stochastic and can
rapidly trigger central bursts by bringing large amounts of
gas to the bottom of the potential well. It is also unclear
what the relation would be between these cold flows and the
central concentration of mass (luminous and dark). Three
galaxies (NGC~1705, NGC~6789, and UGC~9128), however, \textit{seem}
very isolated and show relatively regular optical morphologies
down to $\mu \simeq 26$ R~mag~arcsec$^{-2}$. If the regular
optical morphologies of these three galaxies are confirmed
by deeper optical images, they may represent cases of cosmological
gas accretion in the Local Universe.

\section{Conclusions}

We investigated the distribution and kinematics of the \hi gas
in the outer regions of nearby starburst dwarf galaxies, using both
new and archival data. We considered 18 starburst dwarfs that have
been resolved into single stars by HST observations, providing
their recent SFHs and starburst timescales. Our main results can
be summarized as follows.
\begin{enumerate}
 \item Starburst dwarfs display a broad range of \hi morphologies.
 Several galaxies show heavily disturbed \hi morphologies
characterized by large-scale asymmetries, long filaments, and/or
strong offsets between the stellar and \hi distributions, whereas
other galaxies show only minor asymmetries in the outer regions.
 \item We defined the parameter $A$ to quantify the large-scale \hi
asymmetry in the outer regions and measured it for our sample of
starburst dwarfs and a control sample of typical dwarf irregulars,
drawn from the VLA-ANGST survey. We found that starburst dwarfs
generally have higher values of $A$ than typical irregulars,
suggesting that some external mechanism triggered the starburst.
\item We compared the values of $A$ with the starburst properties.
We found that galaxies hosting a ``young'' burst ($\lesssim$100 Myr)
typically have more asymmetric \hi morphologies than galaxies hosting
an ``old'' one ($\gtrsim$100 Myr), further indicating that there is
a close link between the outer, disturbed \hi distribution and the
central, recent star-formation. Galaxies hosting an ``old'' burst
likely had enough time to partially regularize their outer \hi
distribution, since the ``age'' of the burst ($\sim$500 Myr) is
comparable with the orbital time in the outer parts.
\item We investigated the nearby environment of the galaxies in our
sample. Most of them have a potential perturber at a projected
distance $\lesssim 200$~kpc, thus the hypothesis of a past interaction 
cannot be excluded. Three galaxies (NGC~2366, NGC~4449, and I~Zw~18)
are probably undergoing a minor merger. Another three objects (NGC~1705,
NGC~6789, and UGC~9128), instead, seem very isolated and show regular
optical morphologies down to $\mu \simeq 26$ R mag~arcsec$^{-2}$, thus
they \textit{may} represent cases of cold gas accretion in the nearby
Universe.
\end{enumerate}

\section*{Acknowledgements}
We are grateful to Renzo Sancisi for sharing his valuable insights with
us. We thank Ed Elson, Deidre Hunter, and Angel R. L{\'o}pez-S{\'a}nchez
for providing us with the \hi data of NGC~1705, NGC~4449, and NGC~5253,
respectively. We also thanks the members of the WHISP, THINGS,
LITTLE-THINGS, and VLA-ANGST projects for having made the \hi data
publicly available. FL acknowledges the Ubbo Emmius bursary program
of the University of Groningen and the Leids Kerkhoven-Bosscha Fund.
FF acknowledges financial support from PRIN MIUR 2010-2011, project
``The Chemical and Dynamical Evolution of the Milky Way and Local
Group Galaxies'', prot. 2010LY5N2T.

\appendix

\section{Estimating the noise in a total \hi map}\label{app:noise}

In their appendix A, \citet{Verheijen2001} describe how to calculate the
noise in a total \hi map obtained using a mask on an hanning-tapered
datacube, in which all the channel maps are kept during the analysis.
Here we derive similar formulas that can be used to construct
signal-to-noise maps in 2 different cases: i) a uniform-tapered datacube,
as it is the case for the WHISP data and our new WRST and VLA observations;
and ii) an online hanning-tapered datacube, in which half of the channel
maps are discarded during the observations, as it is the case for the
THINGS/LITTLE-THINGS data and other archival VLA observations.

\begin{table*}
\centering
\caption{Adding $N$ online hanning-tapered channel maps.}
\begin{tabular}{l|c c c c c c c c c}
Channel    & $U_{i-1}$ & $U_{i}$ & $U_{i+1}$ & $U_{i+2}$ & $U_{i+3}$ & ... & $U_{i+2N-3}$ & $U_{i+2N-2}$ & $U_{i+2N-1}$  \\
\hline
$i$        & 1/4       & 1/2     & 1/4       &           &           &     &              &              &                \\
$i+2$      &           &         & 1/4       & 1/2       & 1/4       &     &              &              &                \\
...        &           &         &           &           & ...       & ... & ...          &              &                \\
$i+2N-2$   &           &         &           &           &           &     & 1/4          & 1/2          & 1/4            \\
\hline
           & 1/4       & 1/2     & 1/2       & 1/2       & 1/2       & ... & 1/2          & 1/2          & 1/4            \\
\end{tabular}
\label{tab:sum}
\end{table*}
\subsection*{Uniform taper}

If the observations are made using a uniform velocity taper, the
noise $\sigma^{u}$ in two channel maps will be independent. When $N$
uniform-tapered channel maps are added at the spatial position $(x,y)$,
the noise $\sigma^{u}_{N}$ in the total \hi map will increase by
a factor $\sqrt{N}$, thus $\sigma^{u}_{N}(x,y) = \sqrt{N(x,y)}
\sigma^{u}$. However, before the channel maps are added to form a total
\hi map, the continuum emission is subtracted, introducing further noise
in the channel maps. Here we assume that the continuum map $C^{u}$ is
constructed by averaging $N_{1}$ and $N_{2}$ line-free channel maps
at the high and low velocity ends of the datacube, respectively.
Thus, one has
\begin{equation}
\begin{split}
 C^{u} = \dfrac{1}{2}\bigg(\dfrac{1}{N_{1}} \sum_{j=1}^{N_{1}} U_{j} + \dfrac{1}{N_{2}} \sum_{j=1}^{N_{2}} U_{j}\bigg),
\end{split}
\end{equation}
and the noise $\sigma^{u}_{C}$ in the continuum map is given by
\begin{equation}
\begin{split}
 \sigma^{u}_{C} = \dfrac{1}{2}\sqrt{\dfrac{1}{N_{1}} + \dfrac{1}{N_{2}}} \sigma^{u}.
\end{split}
\end{equation}
If $U_{i}$ is the value of a pixel in the $i^{th}$ uniform-tapered
channel map, the line-emission $L^{u}_{i}$ is given by $L^{u}_{i} =
U_{i} - C^{u}$ and the noise $\sigma^{lu}_{i}$ in $L^{u}_{i}$ is
given by
\begin{equation}
\begin{split}
\sigma^{lu}_{i} = \sqrt{1 + \dfrac{1}{4}\bigg(\dfrac{1}{N_{1}} + \dfrac{1}{N_{2}}\bigg)} \sigma^{u}.
\end{split}
\end{equation}
When $N$ uniform-tapered and continuum-subtracted channel maps are
added, the signal $L^{u}_{N}$ at a position $(x,y)$ of the total
\hi map is given by
\begin{equation}
 L^{u}_{N}(x,y) = \sum_{j=1}^{N(x,y)} L^{u}_{j} = \sum_{j=1}^{N(x,y)} U_{j} - N(x,y) \times C,
\end{equation}
and the noise $\sigma^{lu}_{N}$ is given by
\begin{equation}
\begin{split}
\sigma^{lu}_{N}(x,y)& = \sqrt{N(x,y) \sigma^{u^{2}} + N(x,y)^{2} \sigma^{u^{2}}_{C}} =\\
                    & = \sqrt{1 + \dfrac{N(x,y)}{4} \bigg(\dfrac{1}{N_{1}} +\dfrac{1}{N_{2}}\bigg)} \sqrt{N(x,y)} \sigma^{u}.
\end{split}
\end{equation}

\subsection*{Online hanning taper}

If the observations are made using an hanning taper, the datacube
is smoothed in velocity and the noise in two adjacent channel maps
is no longer independent. When the online hanning smoothing option
of the VLA is used, half of the channel maps are discarded.
If $U_{i}$ and $O_{i}$ are, respectively, the values of a pixel
in the $i^{th}$ uniform-tapered and online hanning-tapered channel
maps, one has
\begin{equation}
\begin{split}
&O_{i} = \dfrac{1}{4} U_{i-1} + \dfrac{1}{2} U_{i} + \dfrac{1}{4} U_{i+1}, \\
&O_{i+1} = \dfrac{1}{4} U_{i} + \dfrac{1}{2} U_{i+1} + \dfrac{1}{4} U_{i+2}, \\
&O_{i+2} = \dfrac{1}{4} U_{i+1} + \dfrac{1}{2} U_{i+2} + \dfrac{1}{4} U_{i+3},
\end{split}
\end{equation}
and the $i+1^{th}$ channel map is discarded during the observations.
The remaining channel maps $i^{th}$ and $i+2^{th}$ are \textit{not}
independent, because both contain a quarter of the emission $U_{i+1}$.
Thus, when $N$ online hanning-smoothed channel maps are added, the
noise $\sigma^{o}_{N}$ does not increase by a factor $\sqrt{N}$,
but by a factor $\sqrt{N - \frac{3}{4}}\frac{4}{\sqrt{2}\sqrt{6}}$,
as we show in the following. The noise $\sigma^{o}$ in the online
hanning smoothed channel maps is equal to $\frac{\sqrt{6}}{4}
\sigma^{u}$ \citep[see][]{Verheijen2001}. The total signal $O_{N}$
is given by
\begin{equation}
O_{N} = O_{i} + O_{i+2} + O_{i+4} + O_{i+6} + ... + O_{i+2(N-1)}.
\end{equation}
As illustrated in Table \ref{tab:sum}, one has
\begin{equation}
\begin{split}
 O_{N} = \dfrac{1}{4} U_{i-1} + \dfrac{1}{2} U_{i} + \dfrac{1}{2} U_{i+1} + ... + \dfrac{1}{2} U_{i+2N-2} + \dfrac{1}{4} U_{i+2N-1},
\end{split}
\end{equation}
and the noise $\sigma^{o}_{N}$ is given by
\begin{equation}
\begin{split}
\sigma^{o}_{\rm{N}} & = \sqrt{ \left( \dfrac{1}{4} \right)^{2} + \left( \dfrac{1}{2} \right)^{2} (2N-2) 
                    + \left(\dfrac{1}{4}\right)^{2}} \sigma^{u} =\\
&= \dfrac{1}{\sqrt{2}} \sqrt{N - \dfrac{3}{4}} \sigma^{u} = \sqrt{N - \dfrac{3}{4}}\dfrac{4}{\sqrt{2}\sqrt{6}} \sigma^{o}.
\end{split}
\end{equation}
The continuum map $C^{o}$ is now constructed by averaging $N_1$ and $N_2$ line-free
channel maps at the high and low velocity ends of the online hanning-tapered datacube,
thus the noise $\sigma^{o}_{C}$ in $C^{o}$ is given by
\begin{equation}
\begin{split}
\sigma^{o}_{C} & = \dfrac{1}{2\sqrt{2}}\sqrt{\dfrac{1}{N^{2}_{1}}\bigg(N_{1}-\dfrac{3}{4}\bigg) +
\dfrac{1}{N^{2}_{1}}\bigg(N_{2}-\dfrac{3}{4}\bigg)} \sigma^{u} \\
               & \equiv \dfrac{1}{\sqrt{2}} A \sigma^{u} = \dfrac{4}{\sqrt{2}\sqrt{6}} A \sigma^{o}.
\end{split}
\end{equation}
The line-emission in the $i^{th}$ channel map is given by $L^{o}_{i}
= O_{i} - C^{o}$, thus the noise in $L^{o}_{i}$ is given by
\begin{equation}
\begin{split}
\sigma^{lo}_{i} = \sqrt{1 + \dfrac{4}{3} A^{2}} \sigma^{o}.
\end{split}
\end{equation}
When $N$ online hanning-tapered and continuum-subtracted channel maps are added,
the signal $L^{o}_{N}$ at a position $(x,y)$ of the total \hi maps is given by
\begin{equation}
\begin{split}
L^{o}_{N}(x,y) &= \dfrac{1}{4}U_{i-1} + \dfrac{1}{2}U_{i} + ... + \dfrac{1}{2} U_{i+2N-2} + \dfrac{1}{4} U_{i+2N-1} + \\
               &- N(x,y) \times C^{o},
\end{split}
\end{equation}
and the noise $\sigma^{lo}_{N}$ at $(x,y)$ is given by
\begin{equation}
\begin{split}
 \sigma^{lo}_{N}(x,y) = \sqrt{N(x,y) - \dfrac{3}{4} + N^{2}(x,y) A^{2}} \dfrac{4}{\sqrt{2}\sqrt{6}}\sigma^{o}.
\end{split}
\end{equation}
Note that this equation differs by a factor $1/\sqrt{2}$ from the one given by \citet{Verheijen2001},
which is valid in the case that all the hanning-tapered channel maps are kept during the data analysis.

\renewcommand\bibname{{References}}
\bibliographystyle{mn2e.bst}
\bibliography{bibliography.bib}

\end{document}